\documentclass[12pt]{article}
\usepackage{epsfig,cite}
\newcommand{\nc}{\newcommand}
\nc{\bb}{\bibitem}
\nc{\be}{\begin{equation}}
\nc{\ee}{\end{equation}}
\nc{\pa}{\partial}
\nc{\ra}{\rightarrow}
\nc{\la}{\leftarrow}
\nc{\etp}{{\eta^\prime}}
\nc{\omg}{\omega}
\nc{\ggam}{\gamma \gamma}
\nc{\bea}{\begin{eqnarray}}
\nc{\eea}{\end{eqnarray}}
\nc{\beas}{\begin{eqnarray*}}
\nc{\eeas}{\end{eqnarray*}}
\nc{\non}{\nonumber}

\begin{document}
\begin{titlepage}
\vbox{~~~ \\
                                   \null \hfill LPNHE 03--07\\
                                  \null \hfill  FERMILAB--PUB-03--166\\
                                   \null\hfill nucl-th/0306078

\title{Anomalous $\eta/\eta^\prime$ Decays~:
\\ The Triangle and Box Anomalies  
   }
\author{
M.~Benayoun$^{(a)}$, P.~David$^{(a)}$, L.~ DelBuono$^{(a)}$ \\
Ph.~Leruste$^{(a)}$, H.~B.~O'Connell$^{(b)}$ \\
\small{$^{(a)}$ LPNHE Paris VI/VII, F-75252 Paris, France }\\
\small{$^{(b)}$ Fermilab, PO Box 500 MS 109, Batavia IL 60510, USA.}\\
}
\date{\today}

\maketitle
\begin{abstract}
We examine  the decay modes 
$\eta/\etp\ra \pi^+ \pi^- \gamma$ within the
context of the Hidden Local Symmetry (HLS) Model.
Using numerical information derived in previous fits 
to $VP\gamma$ and $Ve^+e^-$ decay modes in isolation 
and the $\rho$  lineshape determined in a previous fit
to the pion form factor,
we show that all aspects of these decays can be
predicted with fair accuracy. Freeing  some
parameters does not improve the picture.
This is interpreted 
as a strong evidence in favor of the box anomaly in 
the $\eta/\etp$ decays, which occurs at precisely
the level expected. We also construct the set
of equations defining the amplitudes for
$\eta/\etp\ra \pi^+ \pi^- \gamma$ and
$ \eta/\etp \ra \ggam $ at the chiral limit, as
predicted from the anomalous HLS Lagrangian appropriately broken.
This provides a set of four equations depending 
on only one parameter, instead of three for
the traditional set. This is also shown to match
the (two--angle, two--decay--constant)
$\eta-\etp$ mixing scheme recently proposed and 
is also fairly well fulfilled by the data.
The information returned from fits also matches
expectations from previously published fits to 
the $VP\gamma$ decay modes in isolation.
\end{abstract}
}
\end{titlepage}

\section{Introduction}
\label{introduction}
\indent \indent
Interactions and decays of light mesons fit well within the
framework of Chiral Perturbation Theory (ChPT) \cite{GL}.
Strictly speaking, the ChPT framework applies to the octet
members of the pseudoscalar sector ($\pi,K,\eta_8$) which
behave as Goldstone bosons whose masses vanish at the chiral limit.
Relying on the large $N_c$ limit of QCD, an extended  ChPT 
framework (EChPT) has been defined \cite{Kaiser0,Kaiser} including the 
singlet $\eta_0$ state which keeps a non--zero mass at the chiral 
limit, but this vanishes in the large $N_c$ limit.  
On the other hand, the decays $\pi^0/\eta/\etp \ra \ggam$,
are understood as proceeding from the so--called triangle anomaly.  
These are accounted for by means of the Wess--Zumino--Witten
(WZW) Lagrangian \cite{WZ,Witten}  which is also normally
incorporated into the ChPT Lagrangian \cite{Kaiser0,Kaiser}.
 
Other anomalous processes describing the  $(\pi^0 \pi^+ \pi^- \gamma)$
vertex and the decay mode $(\eta \ra \pi^+ \pi^- \gamma)$
have been identified long ago within the context of Current 
Algebra \cite{Teren}~; they are presently referred to as box anomalies. 
Triangle and box anomalies are now derived from the WZW Lagrangian.
The box anomaly part of  the WZW Lagrangian predicts exactly  the values of 
the amplitudes for the couplings  $(\pi^0 \pi^+ \pi^- \gamma)$, 
$(\eta \pi^+ \pi^- \gamma)$ and $(\etp \pi^+ \pi^- \gamma)$ at the chiral 
limit~; however, the momentum dependence of the corresponding amplitudes 
is not predicted and should be modelled. When dealing with experimental data,
this momentum dependence is naturally accounted for by vector
meson contributions and, then, the question becomes 
whether these alone account for the  box anomalies or whether
an additional contact term (possibly simulating high mass resonances)
is needed~; if this contact term (CT) is needed, it
should have a definite value in order to stay consistent
with the rigorous predictions of the WZW Lagrangian. 

Therefore, from an experimental point of view, the question
of the relevance of the box anomaly phenomenon turns out to check the need
for a well--defined contact term besides the usual resonant contributions. 
This question is still awaiting a definite and unambiguous signature. 

In its simplest figure, the problem of the relevance of the box anomaly
phenomenon is adressed in the coupling $(\pi^0 \pi^+ \pi^- \gamma)$.
The relevance of a possible contact term  beside vector meson 
exchanges has been examined. A value for this coupling has been
extracted from experimental data 
\cite{Antipov} and found close to expectations\footnote{See, also, 
the discussion in \cite{Holstein1}.} (only $2 \sigma$ apart).

A cleaner environment could be provided by the decay modes
$\eta/\eta^\prime \ra \pi^+ \pi^- \gamma$ which are also
accounted for in the WZW Lagrangian. Several pieces of information
are available~: the partial widths \cite{PDG02} are known
with an accuracy of the order 10\% cross--checked
by several means, the $\eta$ spectrum as function of 
the photon momentum has been measured long ago
\cite{Gormley,Layter} and provides useful information.
Finally, measurements of the $\etp$ spectrum
as function of the dipion invariant--mass have been
performed twelve times, with various levels of precision,
and the corresponding data are published 
as papers from several Collaborations
\cite{TASSO,ARGUS,TPC2g,MARKIIa,LeptonF,WA76}
or are available as PhD theses
\cite{Feindt,MARKIIc,McLean,Peters}.
The latest measurements have been performed recently
by CERN Collaborations \cite{Abele,L3}. 

This already represents a large amount of information 
covering all aspects of these decays. This should
allow a reasonably well founded analysis in a search
for the box anomaly in the $\eta/\etp$ system. 

\vspace{0.5cm}

As stated above, predictions on box anomalies are given at the
chiral point and $\eta/\etp$  spectra  clearly extend
in  regions where accounting for the $\rho$  exchange 
cannot be avoided in order to match experimental information. 
The magnitude of the $\eta/\etp \ra \pi^+ \pi^- \gamma$ 
partial  widths should also be influenced by the $\rho$  exchange.
Stated otherwise, including  momentum (invariant--mass)
dependence within their spectra is essential and should
be done in a consistent way in order to extract reliably 
information on the box  anomaly from data.

Indeed, in these decays, two contributions are a priori competing~:
a  contact  term and a 
(dominant) resonant one --the $\rho$ exchange-- with 
contributions of (sometimes) very different magnitudes. 
In order to detect reliably the former, the latter
has to be known with enough accuracy and, possibly, 
fixed. The sharing in the anomalous amplitudes at the chiral 
limit between the contact term and the resonant term
might also have to be understood unambiguously.

Therefore, a global framework is needed where 
the vector meson degrees of freedom are explicitly
accounted for  together with pseudoscalar mesons
and the contact terms. Several such frameworks implementing
vector dominance (VMD) in effective Lagrangians have been 
defined~: Resonance Chiral Perturbation Theory
\cite{GL,Ecker}, massive Yang--Mills fields
\cite{Schechter,Jain,Meissner}, Hidden Local Symmetry (HLS) 
Model \cite{HLS,FKTUY}. It was soon shown \cite{Ecker2}
that all these approaches were physically equivalent.
For convenience, we work within the HLS Model context.

\vspace{0.5cm}

A second issue makes the difference between the $\pi^0$
box anomaly and the $\eta/\etp$ ones. The former is
practically insensitive to symmetry breaking effects
(Isospin Symmetry breaking is a small effect),
the latter however sharply depends on how SU(3)
Symmetry and Nonet Symmetry breakings really take place.
Therefore a reliable breaking scheme of the $\eta/\etp$ 
sector should also be defined and checked in the triangle
and box anomaly sectors. It should also be validated
in all processes where it has to apply,  like 
$V \ra P \gamma$ and $P  \ra V \gamma$ decays. One has already 
noted some confusion \cite{chpt} in the meaning of the 
decay constants entering the amplitudes for the 
$\eta$ and $\etp$ decays to two photons. 

If one limits oneself to collecting some VMD term for the $\rho$ 
contribution (even if motivated) and simply adds it with a phase 
space term to be fit, one can be led to ambiguities  
\cite{Ben0,Ben1} when solving the Chanowitz equations
\cite{Chanowitz} which represent the traditional way
of describing the $\eta/\etp$ mixing (see also \cite{Gilman}). 
Along the same line, if the breaking scheme
generally used \cite{Ben0,Chanowitz,Gilman,Holstein2,Holstein3}
happens to be inappropriate in order to describe
the $\eta/\etp$ system, extracting the box anomaly
constant values from data becomes hazardous. 

A scheme for implementing SU(3) symmetry 
breaking in the full HLS Lagrangian has been already 
defined \cite{BKY,Heath1}. This scheme, referred to as BKY has 
been proved \cite{chpt} to meet all (E)ChPT requirements and
allows a  successful account of a very large
set of experimental data \cite{rad,mixing}.
A brief global account of the full breaking scheme
we advocate is summarized in Appendix A to \cite{ff1}.
The non--anomalous sector has been used in pion form
factor studies providing also consistent results 
\cite{Rho0,ff1}.

Therefore, in this paper, we intend to extend the realm 
of the broken HLS model
by studying the decays $\eta/\etp \ra \pi^+ \pi^- \gamma$.
The behaviour of the model can then be examined
in a context where the box anomaly phenomenon is expected
to be present. One can hope extracting unambiguously the
information about the relevance of this phenomenon from
experimental data.

The paper is organized as follows~; in Section 
\ref{radiative_decays} we recall the traditional 
expressions of the decay amplitudes at the chiral
limit for $\eta/\etp \ra \ggam$ and 
$\eta/\etp \ra \pi^+ \pi^- \gamma$.
In Section \ref{HLS_FKTUY}, we outline the
derivation of the full anomalous sector of the HLS Model,
mostly refering to the basic paper \cite{FKTUY}. In Section
\ref{breaking} we recall shortly how the BKY breaking of
SU(3)  Symmetry and the breaking 
of Nonet Symmetry has been performed and tested.
In Section \ref{two_photon}, we recall the result of applying
this to $\eta/\etp \ra \ggam$ as it provides an unconventional
set of expressions for the amplitudes at the chiral limit.

In Section \ref{box}, we develop the structure predicted
for the $\eta/\etp \ra \pi^+ \pi^- \gamma$ decay modes
by the broken HLS Model. We first show that  
the BKY breaking scheme provides also unconventional
expressions for the box anomaly amplitudes at the chiral limit.
We also show that all information related with these decay modes 
(parameters and $\rho$ meson lineshape) have been already derived 
numerically and functionally in other sectors of the 
low energy phenomenology. It thus follows that all properties  
of the $\eta/\etp \ra \pi^+ \pi^- \gamma$ decay modes
can be predicted without any numerical or functional
freedom. In Section \ref{prediction}, we examine the 
predictions of this model for the $\eta/\etp \ra \pi^+ \pi^- 
\gamma$ partial widths and for their dipion invariant--mass 
spectra. 

After reviewing shortly the status of the
available experimental data on this subject in Section \ref{data},
we devote Section \ref{shapes} to comparing the predicted 
lineshapes with the published experimental spectra.
Section \ref{globalfit} is devoted to performing
a global fit of the shape and yield information for  
the $\eta/\etp \ra \pi^+ \pi^- \gamma$ modes in order
to check precisely the relevance of the numerical
parameters which were all fixed from analysis of
other independent data sets. In Section
\ref{anomalies}, we propose, for comparison,
fits of the anomalous amplitudes at the chiral
limit, under various conditions and show
that the one (instead of three, usually)
parameter dependence of these gets a strong support 
from data. 

Finally, Section \ref{conclusion}  is devoted to a summary
of the results obtained and to conclusions.

\section{Radiative Decays of Neutral Pseudoscalar Mesons}
\label{radiative_decays}
\indent \indent 

Some interactions (or decay modes) of neutral pseudoscalar mesons 
($P=\pi^0,\eta,\eta^\prime$) are described by matrix elements having 
the wrong parity and are called anomalous.
Anomalous interactions were treated by Wess and Zumino \cite{WZ} and 
then expounded upon by Witten \cite{Witten}~; they are
given by the anomalous action, which we shall refer to as
$\Gamma_{\rm WZW}$. For the purpose of this paper, 
two pieces\footnote{Here and in the following, we denote by $V$ the (massive) 
vector field matrix, by $A$ the electromagnetic field and by $P$ 
the pseudoscalar field matrix. The matrix normalization we use for these
have defined in \cite{HLS,Heath1,chpt}~; our normalization for the
SU(3) flavor matrices differs from those in \cite{Holstein1} by a factor of 
2~: $T^a_{Holstein} =  T_{Witten} =2 ~T^a_{HLS}$. Moreover, we use 
without distinction $VP\gamma$ and $AVP$ to name the corresponding coupling.
} from $\Gamma_{\rm WZW}$ are relevant~:

\be
\begin{array}{ll}
{\cal L}_{\gamma \gamma P} =&  \displaystyle
-\frac{N_c e^2}{4 \pi^2 f_{\pi}} 
\epsilon^{\mu \nu \rho \sigma}
\partial_{\mu}A_{\nu} \partial_{\rho}A_{\sigma} {\rm Tr} [Q^2P] \\[0.5cm]

{\cal L}_{\gamma P P P} =&   \displaystyle
-\frac{ieN_c}{3 \pi^2 f_\pi^3}
\epsilon^{\mu \nu \rho \sigma}
A_{\mu}  {\rm Tr} [Q \partial_{\nu}P \partial_{\rho}P\partial_{\sigma}P]
\end{array} 
\label{eq1}
\ee
\noindent with $e^2 =4 \pi \alpha$, and $f_\pi=92.42$ MeV~; 
$Q$ is the quark charge matrix given by{\footnote{There is an intimate connection
between the charge of quarks and the value of $N_c$ in the anomalous action
\cite{Rudaz,Abbas,Bar}~;  $Q=$ Diag(2/3,--1/3,--1/3) if $N_c=3$..}}
$A$ is the electromagnetic field and $P$  is the {\it bare} pseudoscalar field matrix. 
From there, amplitude intensities at the chiral point can be derived.

The first piece ${\cal L}_{\gamma \gamma P}$ describes the decays
$\pi^0/\eta/\eta^\prime \ra \gamma \gamma$. The second piece
${\cal L}_{\gamma P P P}$ contains an interaction term
$\gamma\ra\pi^+\pi^0\pi^-$ briefly discussed in the Introduction. 
This last piece contains also terms which account for the 
anomalous decay modes $\eta/\eta^\prime \ra \pi^+ \pi^- \gamma$. 

Without introducing symmetry breaking effects, the Lagrangian pieces
in Eq. (\ref{eq1}) can give reliable predictions for processes
involving only pions. In order to deal with  interactions involving
$\eta$ or $\etp$ mesons, one has to feed these Lagrangians with
SU(3) and Nonet Symmetry breaking mechanisms. Usually these 
breaking mechanisms are considered to arise from the naive replacement 
of the pseudoscalar decay constants 
\cite{Chanowitz,Gilman,Holstein1,Holstein2,Hajuj}. 
Using obvious notations, the amplitudes at the chiral point 
derived from Eqs. (\ref{eq1}) can be written~:

\be
\begin{array}{ll}
T(X \ra \ggam)& = B_X(0) ~\epsilon^{\mu \nu \rho \sigma}
\epsilon_\mu \epsilon_\nu^\prime k_\rho k_\rho^\prime \\[0.5cm]
T(X \ra \pi^+ \pi^-\gamma)& = E_X(0) ~ \epsilon^{\mu \nu \rho \sigma}
\epsilon_\mu k_\nu p_{\rho}^+ p_{\sigma}^-
\end{array} 
\label{eq2}
\ee

\noindent ($X=\eta,\eta^\prime$), where the  coefficients are,
assuming $N_c=3$~: 

\be
\begin{array}{rl}
B_{\eta}(0) &= - \displaystyle \frac{\alpha}{\pi \sqrt{3}}
\left[ \displaystyle \frac {\cos{\theta_{P}}}{f_8} - 2 \sqrt{2} 
\displaystyle \frac {\sin{\theta_{P}}}{f_0}
\right] \\[0.5cm]
B_{\eta^\prime}(0) &= - \displaystyle \frac{\alpha}{\pi \sqrt{3}}
\left[ \frac {\displaystyle \sin{\theta_{P}}}{f_8} + 2 \sqrt{2} 
\displaystyle \frac{\cos{\theta_{P}}}{f_0}
\right]\\[0.5cm]
E_{\eta}(0) &= - \displaystyle \frac{e}{4 \pi^2  \sqrt{3}} \frac{1}{f^2_{\pi}}
\left[ \displaystyle \frac {\cos{\theta_{P}}}{f_8} - \sqrt{2}
\displaystyle \frac{\sin{\theta_{P}}}{f_0}
\right]\\[0.5cm]
E_{\eta^\prime}(0) &=  - \displaystyle \frac{e}{4 \pi^2  \sqrt{3}}
\frac{1}{f^2_{\pi}}
\left[ \displaystyle \frac {\sin{\theta_{P}}}{f_8} +  \sqrt{2}
\displaystyle \frac{\cos{\theta_{P}}}{f_0}
\right]
\end{array} 
\label{eq3}
\ee

\noindent using the traditional one--angle mixing scheme. The procedure is 
thus obvious~: one replaces one power of $f_\pi$ by the octet ($f_8$) or 
singlet ($f_0$) decay constant understood under their customary definitions 
in (Extended) ChPT. In the following, we refer to $X \ra \ggam$ and 
$X \ra \pi^+ \pi^-\gamma$ as triangle and box anomalies.

\vspace{0.5cm}

This implies several assumptions which are traditionally made
in an implicit way \cite{Holstein1,Holstein2}~:
\begin{itemize}
\item the decay constant $f_8$ and $f_0$ are the (usual) decay constants
of ChPT defined from current expectation values~: 
 $\langle0 | J_{\mu}^8|\eta_8(q)\rangle=i f_8 q_{\mu}$ and
 $\langle0|J_{\mu}^0|\eta_0(q)\rangle=i f_0 q_{\mu}$,
\item SU(3) and Nonet Symmetry Breakings act in exactly the same
way for the triangle and box anomalies.
\end{itemize}

These equations have been used in several ways and they underlie
decades of phenomenological work on the $\eta/\etp$ mixing. 
For instance, Refs. \cite{Gilman,Holstein1,Holstein2} consider 
the two--photon decay widths of the 
$\eta$ and $\eta^\prime$ mesons and the ratio $f_8/f_\pi\simeq 1.3$
from ChPT \cite{GL} to derive $\theta_P \simeq -20^\circ$ and
$f_0/f_\pi \simeq 1.04$~; this meets ChPT expectations \cite{Kaiser0,Kaiser} if
one identifies $\theta_P$ with the presently named $\theta_8$.
Comparable results \cite{Ben1,Holstein3} are derived by using the 
four equations above, after extracting the box anomaly constants from the dipion 
mass spectra in $\eta/\eta^\prime \ra \pi^+ \pi^-\gamma$ decays. 
The third Eq. (\ref{eq3}) has also been used with accepted parameter values
($f_8/f_\pi = 1.25$, $f_0/f_\pi = 1.04$ and $\theta_P=-20.6^\circ$)
inside the HLS Model to derive a successfull description of
$\eta  \ra \pi^+ \pi^-\gamma$ in isolation \cite{Picciotto}.

The validity of the first two Eqs. (\ref{eq3}) has been recently
addressed and consistency of these with the $\eta/\etp$ breaking 
scheme derived from EChPT \cite{Kaiser0,Kaiser} has been 
found doubtfull \cite{chpt,Feldmann1,Feldmann2}.

There is no currently known examination of the last  two 
Eqs. (\ref{eq3})~; however, some remarks  on the renormalization
of the WZW box term \cite{Heath1} tend to indicate that these 
are also doubtful. Therefore, the phenomenological results derived 
from using Eqs. (\ref{eq3}) have to be reexamined in a consistent
framework.

\section{The Anomalous Sector in the HLS Model}
\label{HLS_FKTUY}

\indent \indent

The HLS Model  originally describes the $\gamma - V$ transitions,
all couplings of the kind $VPP$ and possibly $APP$, if the specific
parameter $a$ of the HLS model \cite{HLS} is not fixed
in order to recover the traditional VMD formulation ($a = 2$).
In this framework, the main decay mode $\omega\ra\pi^+\pi^0\pi^-$
of the $\omg$ meson is, for instance, absent  as clear from the
explicit expression of the HLS Lagrangian \cite{Heath1}.

As seen above, anomalous interactions involving pseudoscalar mesons and
photons are contained in $\Gamma_{\rm WZW}$ \cite{WZ,Witten}.
These terms were included in the Hidden
Local Symmetry Lagrangian by Fujiwara {\it et al.} \cite{FKTUY}, along
with anomalous vector meson ($V$) interactions in such a way that the low energy
anomalous processes (in the chiral limit where $m_\pi=0$)
$\gamma\ra\pi^+\pi^0\pi^-$ and $\pi^0\ra\gamma\gamma$
are solely predicted by $\Gamma_{\rm WZW}$. The construction of this
HLS anomalous Lagrangian, originally performed in \cite{FKTUY}, 
is discussed in detail in several excellent
reviews \cite{Meissner,BKY}. Here, we limit ourselves to a brief 
outline of its derivation, pointing out the motivation for
some important assumptions. The anomalous action has the form~:
\be
\begin{array}{ll}
\Gamma&=\Gamma_{\rm WZW}+\Gamma_{\rm FKTUY}\\[0.5cm]
\Gamma_{\rm FKTUY}&=\sum_{i=1}^{4}c_i~\int d^4x\;{\cal L}_i
\end{array}
\label{eq4}
\ee
where the $c_i$ are entirely arbitrary constants. The Lagrangians ${\cal L}_i$
are given in \cite{FKTUY} and each of them contains $APPP$ and $AAP$ pieces which 
would contribute to the anomalous decays, but are cancelled by $APV$ terms. 
These Lagrangians contain also $VPPP$ and $VVP$ pieces \cite{FKTUY}.  

In view of extending the assumption of dominance of vector mesons (VMD)
to the anomalous sector, it was first shown that  a set of $c_i$ in 
Eq. (\ref{eq4}) can be defined in such a way that $\pi^0 \ra \ggam$ occurs 
solely through $\pi^0 \ra \omg \rho^0$ followed by the (non--anomalous) 
transitions $\omg \ra \gamma$ and $\rho^0 \ra \gamma$. The $\pi^0$ width 
thus derived is identical to the Current Algebra prediction reproduced by
 ${\cal L}_{\gamma \gamma P}$ defined in the previous Section.

It was also shown  \cite{FKTUY} that complete vector dominance can be 
achieved, where the direct term $APPP$ is converted to $VPPP$,
with some other set of $c_i$. In this framework, the decay
$\omega\ra\pi^+\pi^0\pi^-$ occurs through $\omega\ra\pi^0 \rho^0$
(the $VVP$ term) followed by $\rho^0 \ra \pi^+ \pi^-$ {\it} and through the 
contact term ($VPPP$) which gives a direct contribution 
$\omega\ra\pi^+\pi^0\pi^-$. However, it happens that 
the $VVP$ contribution (which provides alone the correct $\omg$ 
partial width) is numerically reduced in a significant way by
the contact term ($VPPP$). In view of this, \cite{FKTUY} proposes
another set of $c_i$ which provides an anomalous  
effective Lagrangian containing only a $VVP$ term and, besides, 
the standard WZW term $APPP$ in the following combination~:

\be 
{\cal L}^{\rm FKUTY}=-\frac{3g^2}{4\pi^2}\epsilon^{\mu\nu\rho\sigma}
{\rm Tr}[\pa_\mu V_\nu\pa_\rho V_\sigma P]-\frac{1}{2}{\cal L}_{\gamma PPP}~~,
\label{eq5}
\ee   
\noindent where ${\cal L}_{\gamma PPP}$ is defined in Eqs. (\ref{eq1}).
One should note that the normalization affecting the WZW part
of this Lagrangian is a pure prediction of the HLS Model
based on a definite extension of the VMD concept to anomalous
processes.

Focussing on decays like $\eta/\etp \ra \pi^+ \pi^- \gamma$,
one readily sees from this expression that, in order to recover
the behaviour expected from ${\cal L}_{\gamma PPP}$
alone, these two terms should contribute to the box anomaly
({\it i.e.} the full amplitude at the chiral limit) in the following ratio~:

$$  {\rm VMD} ~~: ~~ {\rm CT}~~  = ~~ -3 ~~ : ~~ 1 ~~~~, $$
\noindent at the chiral limit~; ``VMD'' names here the contribution 
generated by the first term in Eq. (\ref{eq5}) 
and ``CT'' (contact term) those generated from the second term. Thus, the ``VMD'' 
contribution, generated by the triangle anomaly generalized to $VVP$ 
couplings, is predicted dominant at the chiral limit.

Within this framework, the main $\omg$ decay mode proceeds only from
$\omg \ra \rho^0 \pi^0$ followed by $\rho^0 \ra \pi^+ \pi^-$ and
$\phi \ra\pi^+ \pi^-\pi^0$ proceeds solely from $\omg-\phi$ mixing.
The experimental situation concerning the  decay mode $\phi \ra\pi^+ \pi^-\pi^0$
is conflicting. Indeed, a recent result from the SND Collaboration 
\cite{SND} claims for no significant evidence in favor of a contact
$\phi \ra\pi^+ \pi^-\pi^0$ term in their $e^+e^- \ra \pi^+ \pi^-\pi^0$ 
data and provides a new upper bound much more stringent than
previous ones \cite{PDG02}~;  however, using their own data
on the same physical process,  the KLOE Collaboration \cite{KLOE}
claims that a significant contact term is present in their data.
Actually, as there is currently no available analysis performed
using consistenly a full VVP and VPPP Lagrangian or a Lagrangian
like in Eq. (\ref{eq5}), no founded conclusion can really be drawn.

Processes like $\pi^0 / \eta /\eta^\prime \ra \ggam$ occur solely
through $\pi^0 /\eta /\eta^\prime \ra V V^\prime$ followed by
$V,V^\prime \ra \gamma$. However, transitions like $\gamma\ra\pi^+\pi^0\pi^-$
or decays like $\eta/\eta^\prime \ra \pi^+\pi^-\gamma$ receive
contributions from the contact term and from the $VVP$ term (essentially
through $\rho$ meson exchange).

The $VVP$ piece of this effective Lagrangian has been used successfully in 
several recent studies \cite{rad,mixing,chpt,su2} and proved to predict
(after implementing appropriate symmetry breaking mechanisms) up to 26
physics information with a number of free independent parameters ranging 
from 6 to 9 (when Isospin Symmetry breaking is considered \cite{su2}).

\section{The Extended BKY Symmetry Breaking Scheme}
\label{breaking}

\indent \indent 
The study of SU(3) breaking of the HLS Model has been initiated
by BKY \cite{BKY} who proposed a simple and elegant mechanism. However, 
the procedure was soon understood as breaking also Hermiticity
of the derived Lagrangian, which was clearly an undesired
feature. A slight modification \cite{Heath1} of the original BKY 
procedure was shown to cure this problem and to produce a quite 
acceptable broken Lagrangian (see Eq. (A5) in \cite{Heath1}). 
The way field renormalization has to be performed turns out to
define the renormalized field matrix (denoted $P'$) in 
terms of the bare field matrix (denoted $P$) by~:
\be
P= X_A^{-1/2} P^\prime X_A^{-1/2}~~~,
\label{eq6}
\ee

\noindent where the breaking matrix is $X_A={\rm Diag}(1,1,z)$,
with $z=[f_K/f_\pi]^2$. 

As such, the (original) BKY breaking scheme can only address 
a limited amount of physics processes, as all information 
related with the $\eta$ meson can only be treated crudely and 
the properties of the $\etp$ meson cannot be addressed.

\vspace{0.5cm}

In order to address physics information about
the  $\eta/\etp$ system appropriately, the singlet sector has 
first to be introduced in the original HLS Lagrangian.
This has been done by using \cite{Heath1} the U(3)
symmetric field matrix $P=P_8 + P_0$ instead of only
$P=P_8$ when constructing the Lagrangian. This is
found to provide  the HLS Lagrangian with the canonical
kinetic energy of the singlet state ($\eta_0$ field) while
this does not modify the interaction Lagrangian \cite{Heath1}
by adding $\eta_0$--dependent pieces.

The step further is to break the U$_A(1)$ symmetry by
introducing determinant terms \cite{tH} into the
effective Lagrangian which becomes \cite{chpt}~:
\be
\displaystyle
{\cal L}={\cal L}_{HLS} - \frac{1}{2} \mu^2 \eta_0^2
+ \frac{1}{2} \lambda \partial_\mu\eta_0\partial^\mu\eta_0
\label{eq7}
\ee

By means of this (modified) BKY breaking scheme, the
HLS Lagrangian can now address the $\eta/\etp$ system 
with a complete scheme of symmetry breaking (SU(3) and Nonet 
Symmetries). A $\eta_0$ mass term is generated and
the kinetic energy term of the Lagrangian is modified
in a  non--canonical way, which implies that
a field transformation to renormalized fields has to be
performed. This can be done through the two--step 
renormalization procedure defined in \cite{chpt} and 
outlined in the Appendix. This transformation  is well
approximated \cite{chpt} by~:
\be
P= X_A^{-1/2} [P^\prime_8 + xP^\prime_0] X_A^{-1/2}~~~;
\label{eq8}
\ee
\noindent this has been shown to differ \cite{chpt} from the 
exact field transformation only by terms of second order
in the breaking parameters ($[z-1]$, $[x-1]$).  

This transformation \footnote{The motivation behind
this postulate was that weighting differently
the singlet and octet parts
of the $P$ and $V$ field matrices allows to derive
the most general parametrization  \cite{odonnell} 
of the $VP\gamma$ coupling constants consistent with 
only SU(3) symmetry in the vector and pseudoscalar 
sectors, while the corresponding U(3) symmetries 
are both broken.} was postulated (or fortunately anticipated) 
in \cite{rad} in order to study the full set of $AVP$ 
radiative decays, especially the modes involving 
the $\eta/\etp$ mesons. Using this transformation
has provided a fairly good description of the whole
physics accessible to this broken  Lagrangian
\cite{rad,chpt,mixing,su2} with only a 
small number of parameters, as already noted.

Departures from Nonet Symmetry were observed \cite{rad} by 
extracting from data  $x = 0.90 \pm 0.02$, significantly different 
from 1. One may rise the question of the correspondance between
$x$ in Eq. (\ref{eq8}) and the basic Nonet Symmetry breaking parameter 
$\lambda$  of the Lagrangian in Eq. (\ref{eq7}) -- which is
\cite{chpt}  nothing but the $\Lambda_1$ parameter of 
\cite{Kaiser0,Kaiser}. One can write~:
\be
\displaystyle
x=\frac{1}{\sqrt{1+ h \lambda}}~~~~, ~~h=1+{\cal O}(z-1)~~~.
\label{eq9}
\ee

Indeed, as shown in the Appendix, $x$ absorbs a small
influence of SU(3) symmetry breaking (about 5\% of its fitted value).
 
\vspace{0.5cm}

Comparison of all results of this broken HLS Lagrangian, especially decay 
constants and mixing angles, with the available (E)ChPT estimates of
the same parameters \cite{Kaiser0,Kaiser,Feldmann1}  was done and appeared
also fully satisfactory. It is worth remarking that the HLS 
phenomenology was yielding an estimate for the (E)ChPT mixing angle
$\theta_0$ much smaller in magnitude than the (E)ChPT leading order
estimate ($-0.05^\circ \pm 0.99^\circ$ in contrast with $\simeq -4^\circ$),
but in fair correspondence with a more recent EChPT next--to--leading order 
calculation \cite{Holstein4} which yields $\theta_0=[-2.5^\circ,+0.5^\circ]$.

It is also worth remarking that the (full) breaking scheme
just outined anticipated \cite{rad} the branching fraction for 
$\phi \ra \etp \gamma$ with a value twice smaller
than its contemporary measurement \cite{CMD2_etp}. This predicted 
value coincides with all recent measurements
performed with larger statistics \cite{PDG02}.

\vspace{0.5cm}

The quasi--vanishing of $\theta_0$  has two interesting
consequences. On the one hand, it allows to relate the
traditional wave--function mixing angle with the recently
defined $\theta_8$ mixing angle \cite{Kaiser0,Kaiser} by providing
$\theta_8 \simeq 2 \theta_P$ (fulfilled at a few percent level)~;
the derivation is given in the Appendix for the exact field
transformation.

On the other hand, the condition $\theta_0=0$ relates the Nonet symmetry 
breaking parameter $x$ to $\theta_P$~: 

\begin{equation}
\displaystyle \tan{\theta_P}=\sqrt{2} \frac{(1-z)}{2+z} ~x
\label{eq10} 
\end{equation}

This relation is fulfilled with a high degree
of numerical accuracy  \cite{chpt} and only reflects
that the ChPT mixing angle $\theta_0$ is 
not significantly affected by symmetry breaking effects.
This relation will be somewhat refined
(See Section \ref{anomalies} and the Appendix).

It then follows that, from the three originally free breaking
parameters associated with the pseudoscalar sector ($z$, $x$, $\theta_P$), 
only one remains unconstrained. It could  be either of $x$ or 
$\theta_P$~; however, it will be shown that $x$ might
be prefered.  

We do not go on discussing here symmetry breaking in 
the vector meson sector as it is not in the stream of the present 
paper~; we refer interested readers to \cite{BKY,Heath1,rad,mixing} 
where this is discussed in details.

Some remarks are of relevance. The combined Nonet Symmetry and SU(3) breaking 
scheme of the HLS Lagrangian presented in this Section defines
what we name the Extended BKY breaking scheme. It restores the
relevance of  a one angle mixing scheme for the $\eta/\etp$ system.
However, this does not give any support to the traditional
breaking scheme as expressed by Eqs. (\ref{eq3}). In contrast,
it matches fairly well all expectations of the two--angle, two--coupling
constant mixing scheme  recently derived from
EChPT \cite{Kaiser0,Kaiser,Feldmann1,Feldmann2} at
leading order in the breaking parameters. 

This full breaking scheme is also mathematically 
equivalent to the recently proposed \cite{Feldmann1,Feldmann2}
breaking in the quark flavor basis framework~; it might be prefered  as
a definite concept like Nonet Symmetry breaking, which underlies some 
Lagrangian pieces (${\cal L}_2$) of EChPT, can be implemented more clearly
and tracing its effect in phenomenology is easier.

\section{Two--Photon Decay Widths of the $\eta$ and $\eta^\prime$ mesons}
\label{two_photon}

\indent \indent
The two--photon decay widths of the $\eta$ and $\eta^\prime$ mesons
can be derived easily from the HLS Lagrangian (the $VVP$ part of Eq.
(\ref{eq5})) supplemented by the $V\gamma$ transition amplitudes
of the non--anomalous HLS Lagrangian)
after renormalizing to physical fields by Eq. (\ref{eq8}). Applying
directly the same Eq. (\ref{eq8}) to the WZW Lagrangian 
${\cal L}_{\gamma \gamma P}$ in Eq. (\ref{eq1}) leads exactly
to the same result\footnote{From now on, we assume $N_c=3$.} \cite{chpt}~:

\be
\begin{array}{rl}
G_{\eta}(0) &= -\displaystyle \frac{\alpha}{\pi \sqrt{3} f_{\pi}}
\left [ \frac{5z-2}{3z}\cos{\theta_P}-\sqrt{2} \frac{5z+1}{3z}x \sin{\theta_P} 
\right ],\\[0.5cm]

G_{\eta^\prime}(0) &= -\displaystyle \frac{\alpha}{\pi \sqrt{3} f_{\pi}}
\left [ \frac{5z-2}{3z}\sin{\theta_P} + \sqrt{2} \frac{5z+1}{3z}x 
\cos{\theta_P} \right ]~~.
\end{array}
\label{eq11}
\ee

These expressions compare well with the corresponding quantities 
in Eqs. (\ref{eq3}). However, this correspondence is only formal
as, defining $\overline{f}_8$ and $\overline{f}_0$  by~:

\be
\frac{f_{\pi}}{\overline{f_8}}=\frac{5z-2}{3z}~,\,\,\,\,\,\,
\frac{f_{\pi}}{\overline{f_0}}=\frac{5z+1}{6z}~x~,
\label{eq12}
\ee
\noindent yields $\overline{f}_8=0.82 f_\pi$ (and 
$\overline{f}_0=1.17 f_\pi$), which has little to
do with numerical expectations from ChPT (extended or not).
It was  proved in \cite{chpt} that  these 
are {\it not} the standard  EChPT decay constants.
These can  be derived from our broken Lagrangian,
yielding information which matches \cite{chpt} fairly well 
EChPT expectations  \cite{Kaiser0,Kaiser}.

This proved the basic consistency of breaking scheme
presented in the previous Section 
with EChPT. The formulation given in Eq. (\ref{eq11})
could look, at leading  order, more tractable than present 
standard expressions~; it makes indeed much clearer that the number
of parameters to be determined phenomenologically is limited.

\vspace{0.5cm}

\indent
Therefore, the first basic assumption  which underlies
the understanding of Eqs. (\ref{eq3}) is not fulfilled
by the BKY breaking scheme \cite{BKY,Heath1} and this is independent
of whether Nonet Symmetry is broken. 

Let us remind that Eqs. (\ref{eq11})
give  the two--photon radiative decay widths of 
the $\eta/\eta^\prime$ mesons with good accuracy. These can 
even be predicted by using solely the value 
of $x$ extracted from fit 
to the (independent) set of $AVP$ decay modes of light mesons 
\cite{chpt}. Fixing $z=[f_K/f_\pi]^2$ to its experimental
value and assuming Eq. (\ref{eq10}), Eqs. (\ref{eq11}) become
a constrained system depending on only one parameter
and can be solved providing results consistent
with using, instead, the $AVP$ decay mode information. 

Of course, the mixing angle $\theta_P$ entering Eqs. (\ref{eq11})
 does not  coincide with $\theta_8$ and 
is derived \cite{chpt} as $\theta_P=-10.32^\circ \pm 0.20^\circ$ 
when requiring the constraint Eq. (\ref{eq10}) to hold exactly~;  
the corresponding value for $\theta_8 \simeq -20^\circ$ 
compares well to expectations \cite{GL,Kaiser0,Kaiser}. 
One should note a recent estimate of $\theta_P=-10^\circ \pm 2^\circ$ 
provided by lattice QCD computations \cite{McNeile} which strongly 
supports this phenomenologically extracted value. 

Therefore, the picture represented by Eqs. (\ref{eq11}), which does
not meet traditional expectations \cite{Holstein1,Holstein3,Holstein4,Ben1},
matches quite well all relevant information from ChPT and QCD, and,
last but not least, corresponds to  a satisfactory description
of the whole set of two--body radiative decays of light mesons
\cite{rad,mixing,chpt}.
  
\section{The HLS Model For $\eta/\eta^\prime \ra \pi^+ \pi^- \gamma$ Decay Modes}
\label{box}

\indent \indent
Using the effective Lagrangian in Eq. (\ref{eq5}),
the processes $\eta/\eta^\prime \ra \pi^+ \pi^- \gamma$
receive VMD contributions from the $VVP$ term and CT
contributions from the  ${\cal L}_{\gamma PPP}$ piece.
The purpose of this Section is to examine carefully these
decay modes. These will also lead us to question the
last two (box) anomaly equations Eqs. (\ref{eq3}).

\subsection{Basic Lagrangians}

\indent \indent
Within the HLS Model, the $VVP$ part of  $\eta/\eta^\prime \ra \pi^+ \pi^- \gamma$
involves, beside the $\rho$ meson, the interplay of the $\omg$ and 
$\phi$ mesons to their decay mode to $\pi \pi$ only\footnote{Indeed, the 
non--anomalous HLS Lagrangian, broken or not \cite{Heath1}, contains 
no couplings like $\eta \pi V$ or $\eta^\prime \pi V$.
Therefore, terms like $\eta/\eta^\prime \ra \pi V$ 
followed by $V \ra \pi \gamma$ do not contribute to the decays
under examination.}. 
However, these (isospin violating) couplings are small enough to 
be safely neglected. Additionally, the $\phi$
meson is outside the decay phase space of both $\eta$ and 
$\etp$ mesons, and the accuracy of the data is
far from allowing any $\omg$ effect to be significant or simply 
visible in the $\etp$ dipion invariant--mass spectrum.

It can be shown \cite{mixing}, that the $VP\gamma$ couplings
following from the anomalous sector of HLS model can be derived
from the corresponding ($VP\gamma$) piece of~:
\be
{\cal L}=C \epsilon^{\mu \nu \rho \sigma}
{\rm Tr}[
\partial_\mu(eQA_\nu + g V_\nu)\partial_\rho (eQA_\sigma + g V_\sigma)
X_A^{-1/2}(P^\prime_8 + x P^\prime_0)X_A^{-1/2}
]~~,
\label{eq13}	
\ee
\noindent where $g$ is the universal vector coupling of the 
HLS Model \cite{HLS}. The value for $C=-3/(4 \pi^2 f_\pi)$
is fixed by normalizing the $AAP$ term in Eq. (\ref{eq13})
to the corresponding WZW Lagrangian in Eq. (\ref{eq1}).

This equation could essentially be
considered as a way to postulate VMD for $VVP$ couplings
and it also gives the normalization shown in Eq. (\ref{eq5}). 
Focussing on the piece of Eq. (\ref{eq13}) related 
with neutral pseudoscalar mesons, one gets~:
\be
{\cal L}_{\gamma \rho P^0}= -\frac{eg}{4 \pi^2 f_\pi}
\left [
\frac{1}{2} \pi^0 + \frac{\sqrt{3}}{2} \eta_8 + x \sqrt{\frac{3}{2}} \eta_0
\right ]
\epsilon^{\mu \nu \alpha \beta} \partial_\mu A_\nu \partial_\alpha \rho^0_\beta~~,
\label{eq14}	
\ee
\noindent with obvious notations. The $\rho$ meson decay 
relevant for the present study is driven by~:
\be
{\cal L}_{\rho \pi \pi}= i\frac{ag}{2} \rho^0_\mu
\left [
\pi^-\partial^\mu \pi^+ - \pi^+\partial^\mu\pi^-
\right ]
\label{eq15}	
\ee
\noindent which can be extracted from the non--anomalous
(broken or not) HLS Lagrangian. The parameter $a$ is the specific
HLS parameter expected such that $a=2$ in traditional
formulations of VMD, but has always been fitted in the range 
$a=2.3 \div 2.5$ from radiative and leptonic decays of light mesons 
\cite{rad,mixing} and in pion form factor
studies \cite{Rho0,CMD2}.  We recall that the $\rho$ mass
in the HLS Model is not free and is given by the (extended) KSFR 
relation $m_\rho^2=a g^2 f_\pi^2$ which  does not coincide
with traditional mass values \cite{PDG02} ($\simeq 830$ MeV versus 
$\simeq 775$ MeV)~; these happen to be only a matter of definition 
\cite{ff1}. Finally, the CT contributions are contained 
in the following Lagrangian piece extracted from Eq. (\ref{eq5}) 
--for the normalization-- and Eq. (\ref{eq1})~:
\be
{\cal L}_{\gamma \pi^+ \pi^- P^0}=
- i \frac{e}{8 \pi^2 f_\pi^3}
\epsilon^{\mu \nu \alpha \beta}
A_\mu \partial_\nu \pi^+ \partial_\alpha \pi^-
\left [
\partial_\beta \pi^0 + \frac{1}{\sqrt{3}} \partial_\beta \eta_8 
+ x \sqrt{\frac{2}{3}} \partial_\beta \eta_0
\right ]~~.
\label{eq16}	
\ee

Changing from $\eta_8,\eta_0$ to $\eta,\eta^\prime$ is performed by a 
rotation involving the (wave--function) mixing angle~:
\be
\left[
     \begin{array}{l}
     \displaystyle \eta   \\[0.5cm]
     \displaystyle \eta'  
     \end{array}
\right]
=
\left[
     \begin{array}{lll}
\displaystyle \cos{\theta_P} & -\displaystyle \sin{\theta_P} \\[0.5cm]
\displaystyle \sin{\theta_P} & ~~\displaystyle \cos{\theta_P}
     \end{array}
\right]
\left[
     \begin{array}{ll}
     \eta_8\\[0.5cm]
     \eta_0
     \end{array}
\right]
\label{eq17}
\ee

There is, obviously, no loss of generality in introducing
this definition and, thus, the physical parameter $\theta_P$
which has to be fixed or fitted. 

For all expressions in this Section, the fields which occur are
the renormalized ones. It should already be noted that all basic
Lagrangian pieces involved in the considered $\eta/\eta^\prime$ 
decays do not depend at leading order on the SU(3) breaking
mechanism (the parameter $z=[f_K/f_\pi]^2$ already met), at
least in the limit of Isospin Symmetry where we stand. 
However SU(3) symmetry breaking is hidden inside $\theta_P$
(see Eq.(\ref{eq10})).

\subsection{Amplitudes and Chiral Limit Properties}

\indent \indent 
With the information given just above, it is an easy task
to compute the amplitudes for the $\eta/\eta^\prime$
decays considered. One finds~:
\be
\displaystyle
T(X \ra \pi^+ \pi^- \gamma) = c_X \frac{ie}{8\sqrt{3}\pi^2 f_\pi^3} 
\left[
1 + \frac{3 m_\rho^2}{D_\rho(s)}
\right ]
\epsilon^{\mu \nu \alpha \beta}
 \epsilon_\mu k_\nu p^+_\alpha p^-_\beta~~,
\label{eq18}
\ee

\noindent using obvious notations, with $X=\eta,\eta^\prime$,
$m_\rho^2=a g^2 f_\pi^2$ and $s$ being the dipion invariant--mass.
The $c_X$ are given by~:
\be
\begin{array}{ll}
c_\eta &= [\cos{\theta_P} -x \sqrt{2} \sin{\theta_P}]~~, \\[0.3cm]
c_{\eta^\prime} &= [\sin{\theta_P} +x \sqrt{2} \cos{\theta_P}]~~.
\end{array}
\label{eq19}
\ee

\noindent  Finally $D_\rho(s)$ is the inverse $\rho$ 
propagator which can be written \cite{ff1}~:
\be
 D_\rho(s)=s - m_\rho^2 -\Pi_{\rho \rho}(s)~~,
\label{eq20}
\ee
\noindent in terms of the already defined (KSFR) $\rho$ mass and of 
the $\rho$ self--mass. The occurence of $\Pi_{\rho \rho}(s)$
permits to move the $\rho$ pole off from the physical region
\cite{Pich,ff1}. For the present purpose, one has only to stress 
that the $\rho$ self--mass can be chosen  rigorously such 
that $\Pi_{\rho \rho}(0)=0$ as expected from current 
conservation \cite{O'Connell:1994uc}.

Going to the chiral limit, Eq. (\ref{eq18}) is nothing but
the second Eq. (\ref{eq2}) and has for coefficients~: 
\be
\begin{array}{rl}
E^\prime_\eta (0) & \displaystyle= -\frac{ie}{4\sqrt{3}\pi^2 f_\pi^3}
[\cos{\theta_P} -x \sqrt{2} \sin{\theta_P}] \\[0.5cm]
E^\prime_{\eta^\prime} (0) &\displaystyle = -\frac{ie}{4\sqrt{3}\pi^2 f_\pi^3}
[\sin{\theta_P} +x \sqrt{2} \cos{\theta_P}]~~.
\end{array}
\label{eq21}
\ee

These relations clearly meet expectations of Current Algebra~; they 
could not be derived exactly if $\Pi_{\rho \rho}(0) \neq 0$. 
The single symmetry breaking parameter occuring manifestly 
is $x$ which essentially measures departures from Nonet Symmetry.   

It should be noted that there is no obvious connection between
the way symmetry  breaking occurs for the box anomaly (Eq. (\ref{eq21}))
and for the triangle anomaly (see Eqs. (\ref{eq11})) within
the broken HLS Model. It is worth recalling once more, that 
the symmetry breaking scheme defined in Section \ref{breaking} was 
shown \cite{chpt} to match perfectly all expectations of EChPT 
collected in \cite{Kaiser0,Kaiser,Feldmann1}  and that no 
further piece has been added in order to derive Eqs. (\ref{eq21}).

Therefore, the second assumption which underlies the traditional
way of breaking symmetries for this set of equations (see the discussion
after Eqs. (\ref{eq3})) is also not met by the BKY breaking scheme
\cite{BKY,Heath1,chpt}. 

\vspace{0.5cm}

Eqs. (\ref{eq21}) are also interesting for other aspects~:
In order to recover the values expected for both 
$E^\prime_X(0)$, the VMD ({\it i.e.} resonant) contribution happens 
to be 3 times larger than the contact term (CT) contribution
and carries an opposite sign~; this was expected
when building the anomalous HLS Lagrangian Eq. (\ref{eq5}). 

Eqs. (\ref{eq18}) also show that fitting 
the $\eta/\etp$ invariant--mass spectra with a constant
term interfering with a resonant term is indeed legitimate.   
However, it is also clear that the value
of this constant is {\it not} the value of the full amplitude 
at origin and thus carries only a part of the box anomaly
value.

\vspace{0.5cm}

The triangle and box anomaly expressions in the broken HLS model
are summarized by Eqs. (\ref{eq11}) and (\ref{eq21}). 
We know from previous works \cite{rad,chpt} that experimental data support 
this in the triangle anomaly sector  ($AVP$ and $\eta/\etp \ra \ggam$)~;
the real issue is  to test its validity in 
processes where box anomalies are expected to occur.

\subsection{The $\gamma \pi^+ \pi^- \pi^0$ Amplitude}

\indent \indent Even if outside the main stream of this paper, it is 
interesting to give the amplitude for the 
$\gamma \pi^+ \pi^- \pi^0$ anomalous coupling. Using Eqs. (\ref{eq1}) 
and (\ref{eq14}) above, together with the piece analogous to
Eq. (\ref{eq15}) for $\rho^\pm$ which can be found in \cite{Heath1}, one gets~:
\be
T(\gamma \ra \pi^+ \pi^- \pi^0)=A(s,t,u) ~\epsilon^{\mu \nu \alpha \beta}
 \epsilon_\mu p^+_\nu p^-_\alpha p^0_\beta~~,
\label{eq22}
\ee
\noindent with~:
\be
A(s,t,u)=\displaystyle
\frac{e}{8 \pi^2 f_\pi^3}\left[
1+ \frac{m_\rho^2}{D_\rho(s)}+ \frac{m_\rho^2}{D_\rho(t)}+\frac{m_\rho^2}{D_\rho(u)}
\right ]~~,
\label{eq23}
\ee
\noindent where $m_\rho^2=a g^2 f_\pi^2$ and $D_\rho$ is given by
Eq. (\ref{eq20}) with the appropriate permutation of the argument
$s$ to $t$ and $u$. In this case, the symmetry breaking mechanism
we have defined has no influence. The momentum dependence of this 
expression differs from known ones (recalled by Eqs. (39) and (40)
in \cite{Holstein1}) by its taking into account the $\rho$ self--energy
(see Eq. (\ref{eq20})). It gives the expected
value ($-ie/4\pi^2 f_\pi^3$) at $s=t=u=0$ and is worth to 
be checked on forthcoming experimental data \cite{Jefferson}.
It might also be extracted from $e^+ e^- \ra \pi^+ \pi^- \pi^0$
annihilation  with data covering the low invariant--mass region.

\subsection{Invariant--Mass Spectra And The Box Anomaly}
\label{function}

\indent \indent 
>From the amplitude in Eq (\ref{eq18}), one derives the
decay width~:
\be
\displaystyle \frac{d\Gamma(X \ra \pi^+ \pi^- \gamma)}{d\sqrt{s}} =\displaystyle 
\frac{c_X^2}{36} \frac{\alpha}{[2 \pi f_\pi]^6} 
\left| 1+ \frac{3 m_\rho^2}{D_\rho(s)} \right|^2
k_\gamma^3 q_\pi^3~~,
\label{eq24}
\ee
\noindent for each of the $\eta$ and $\eta^\prime$ mesons. We have
defined $k_\gamma=(m_X^2-s)/2m_X$, the photon momentum
in the $X$ rest frame and $q_\pi =\sqrt{s-4m_\pi^2}/2$, the
pion momentum in the dipion rest frame.

The contact term generated by the 
second piece in Eq. (\ref{eq5}),  is represented in Eq. (\ref{eq24})
by the number 1 inside the modulus squared. On the other hand, as
the normalization of the VMD contribution can be fixed \cite{rad,mixing}
at the appropriate value by only normalizing to the $P \gamma \gamma$
amplitude
in Eq. (\ref{eq13}), checking the effect of this contact term by switching 
on/off this ``1'' in Eq. (\ref{eq24}) is indeed meaningfull. In this way, 
one can address the experimental relevance of Eq. (\ref{eq5}).

\vspace{0.5cm}

\indent
Eq. (\ref{eq24}) is interesting in many regards~:

\begin{itemize}
\item The shape of the invariant--mass spectra depends
 on the $\eta/\eta^\prime$ meson properties only through a kinematical
 factor ($k_\gamma^3$).
Therefore, the shape of the invariant--mass spectra does not
carry any {\it manifest} information on the box anomaly constants $c_X$,
\item The lineshape of the 
invariant--mass spectra in $\eta/\eta^\prime$ decays depends only 
on $\rho$ meson properties. However, 
the way this dependence occurs in $\eta/\eta^\prime$ decays is different  
from the one in the pion form  factor \cite{ff1}, as the dressing of 
the $\rho-\gamma$ transition amplitude $\Pi_{\rho \gamma}(s)$ plays no role
in the $\eta/\etp$ decays, 
\item All information on the value of $c_\eta$ and $c_{\eta^\prime}$
is carried by the partial width itself. It can be algebraically
derived if $D_\rho(s)$ is known reliably from another source.

\end{itemize}

In order to perform a search for the box anomaly, one needs a function
$D_\rho(s)$ accurately determined between the two--pion threshold
and the $\phi$ mass.  In the physical region involved in $\eta/\etp$ decays, all 
coupled channels allowed by the HLS model contribute at  
one--loop order \cite{ff1}. However, except for $\pi^+ \pi^-$, each
provides\footnote{And, to some extent, except also for the $\omg \pi^0$ channel
in the $\etp$ decay~; however, the neglected effect is concentrated 
in the region above 917 MeV, very close to the phase 
space boundary for $\etp$ decay and far beyond in the $\eta$ decay.} 
only logarithms, beside their influence 
on the subtraction polynomial hidden inside the $\rho^0$ self--mass 
$\Pi_{\rho \rho}(s)$. This is their major effect, and thus neglecting these 
loops while still considering a subtraction polynomial of the
appropriate degree is certainly motivated.

 Reduced to  only its coupling to $\pi^+ \pi^-$ (with 
or without accounting for kaon pairs), the $\rho$ propagator used here
contributes to providing a fairly accurate numerical determination 
of the pion form factor both in modulus\cite{ff1,Pich} and in phase 
\cite{ff1} up to the $\phi$ mass.  Therefore,
for the purpose of studying the box anomaly, there no point in 
going beyond contributions from only the non--anomalous HLS Lagrangian. 
In this case, the $\rho$ self--energy can be written~:
\be
\Pi_{\rho \rho}(s)=  \displaystyle
g^2_{\rho\pi\pi} [\ell_\pi(s) +\frac{1}{2z^2} \ell_K(s)]~~,\\[0.3cm]
\label{eq25}
\ee

\noindent where $g_{\rho\pi\pi}=ag/2$, while $z$ has been already 
defined. We have denoted 
$\ell_\pi(s)$ and $\ell_K(s)$ the pion and kaon loops
amputated from their couplings to external legs (we neglect
the mass difference between $K^\pm$ and $K^0$)~; these
are given in closed form in \cite{mixing}. 

These loops should be subtracted at least twice in order to 
make convergent the Dispersion Integrals which define them
as analytic functions~; this gives rise to a first degree polynomial
$P_\rho(s)$ with arbitrary coefficients to be determined by imposing
explicit conditions or by fit.  However, as noted just above, anomalous 
loops force to perform, at least, three subtractions
\cite{mixing,ff1}, which modifies the arbitrary polynomial $P_\rho(s)$ 
to (at least) second degree. It is the reason why $P_\rho(s)$ 
will be assumed of second degree, even if one limits oneself 
to pion and kaon loops. This  does not increase our parameter
freedom, as will be seen shortly. 

\subsection{External Numerical Information}
\label{numbers}

\indent \indent
>From what seen above, the condition $P_\rho(0)=0$ on the subtraction
polynomial is certainly desirable, otherwise the Current Algebra 
expectations could not be derived
exactly~; additionally, this condition ensures masslessness of
the photon after dressing by pion and kaon loops.
It thus remains 2 subtraction constants to be determined
or chosen~; we shall fix them from the fit performed \cite{ff1} 
on the pion form factor  from threshold to the $\phi$ mass.
Denoting $\overline{\Pi}_{\rho\rho}(s)$ the $\rho$ self--energy
subtracted three times \cite{mixing}, we have~:
\be
\Pi_{\rho \rho}(s)=  \displaystyle
\overline{\Pi}_{\rho\rho}(s) + e_1 ~s +e_2 ~s^2~~.\\[0.3cm]
\label{eq26}
\ee

On the other hand, we can also fix the HLS parameters $a$, $g$ 
(and thus $m_\rho$) and $x$ to their values fitted in radiative 
and leptonic decays\footnote{Actually, the values for  $g$ and $x$
are determined almost solely by the $AVP$ radiative decays~;
the value for $a$ is a consequence of these on the $V \ra e^+ e^-$ 
decay modes, but mostly the $\omg$  and $\phi$ leptonic decays.} 
\cite{rad,mixing}. Knowing $z$ and $x$, 
one can derive the value for $\theta_P$ from Eq. (\ref{eq10}).
The values for  $e_1$ and $e_2$ are fixed from an appropriate
fit to the pion form factor \cite{ff1}, where the parameters
$a$, $g$ and $x$ are fixed consistently with $AVP$ and
$(\omg/\phi)  e^+ e^-$ modes.

As we restrict the $\rho$ coupling to only the $\pi^+\pi^-$  
and $K \overline{K}$ channels,  these values for $e_1$ and 
$e_2$ are certainly correlated with the chosen values for 
$a$ and $g$~; they are not affected numerically by the
value for $x$.

\vspace{0.3cm} 

\begin{table}[htb]
\begin{center}
\begin{tabular}{lr}
Parameter & Value   \\
\hline
$a$        & $2.51 \pm 0.03$  \\ [0.3cm]
$g$        & $5.65 \pm 0.02$  \\ [0.3cm]
$x$        & $0.90 \pm 0.02$  \\ [0.3cm]
$z=[f_K/f_\pi]^2$    & $1.51 \pm 0.02$  \\ [0.3cm]
\hline
$e_1$		& $0.222 \pm 0.011$ $~~~~~~~~~~~$\\ [0.3cm]
$e_2$		& $-1.203 \pm 0.017$ GeV$^{-2}$\\ 
\hline
\end{tabular}
\caption{Parameter values for $a$, $g$, $x$ ,$z$
fixed from a global fit to $VP\gamma$ and $V \ra e^+ e^-$ 
decay modes \cite{rad,mixing,su2}~; $e_1$ and $e_2$ are 
fixed from a fit \cite{ff1} to the pion form 
factor including only $\pi \pi$ and $K \overline{K}$
as channels coupling to $\rho$.}
\label{T0}
\end{center}
\end{table}
\indent
The values for these parameters are gathered in Table \ref{T0}.
As these parameters are supposed universal in the realm of
the HLS Model, one can fix their values from fit to data independent
of the $\eta/\eta^\prime \ra \pi^+ \pi^- \gamma$ decay modes.
It is indeed the case for the $VP\gamma$ or the $Ve^+e^-$ decay modes
and for the pion form factor. As commented on above, these fit 
values correspond to a very good fit quality for the corresponding data. 
For instance, they allow to {\it predict} \cite{chpt} the two--photon 
decay widths as recalled in Table \ref{T1}.

\vspace{0.3cm} 

\begin{table}[htb]
\begin{center}
\begin{tabular}{lrrr}
Parameter & PDG 2002 & Prediction  & Significance ($n ~\sigma$) \\
\hline
$\eta \ra \ggam$    (keV)             & $0.46 \pm 0.04$  & $0.46 \pm 0.03$ 
& $0.0 ~\sigma$ \\ [0.3cm]
$\eta^\prime \ra \ggam$   (keV)       & $4.29 \pm 0.15$  & $4.41 \pm 0.23$ 
& $0.4 ~\sigma$ \\
\hline
\end{tabular}
\caption{Partial widths for $\eta/\eta^\prime \ra \ggam$
as predicted using Eq. (\ref{eq11}) with parameter values
as coming from a global (HLS) fit to only $VP\gamma$ 
decay modes of light mesons \cite{chpt}.}
\label{T1}
\end{center}
\end{table}
\indent
Choosing the $\rho$ propagator as it comes out of the HLS fit to 
the pion form factor \cite{ff1} is also fairly legitimate, as
this $\rho$ propagator should also be valid anywhere  
within the HLS framework. As stated above, we  consider 
for clarity only the case where the only open channels are 
$\pi \pi$ and $K \overline{K}$. We have, nevertheless, checked that 
changing to various open channel subsets coupling to the $\rho$ meson 
(as done in \cite{ff1} for the pion form factor), with correspondingly 
changing  $e_1$ and $e_2$ to their fit values, does not produce any 
significant modification to the results presented below. 

To summarize, self--consistency implies that we can fix
all parameters and functions from their most reliable fit
values  and expressions, provided the data set is independent
of the $\eta/\etp$ decays considered here. This independent
data sample covers  $VP\gamma$ and $Ve^+e^-$ couplings 
\cite{rad,mixing,chpt} and the pion form factor \cite{ff1}. 
It then follows that all information related with the box anomalies 
can be {\it predicted} without any parameter freedom.

\section{Predictions For $\eta/\eta^\prime \ra \pi^+ \pi^- \gamma$ Decays}
\label{prediction}

\indent \indent In this Section, we examine the predictions derived
for the  $\eta/\eta^\prime \ra \pi^+ \pi^- \gamma$ decay modes
for the partial widths and dipion invariant--mass distributions.

\subsection{Partial Widths, Experimental Values and Predictions}

\indent \indent Using Eq. (\ref{eq24}) and numerical (and functional)
information given in the previous Subsection, it is easy to check that
we can write~:
\be
\Gamma(X \ra \pi^+ \pi^- \gamma) = A_X c_X^2~~,
\label{eq27}
\ee
\noindent where~:
\be
A_X=\displaystyle \frac{1}{36} \frac{\alpha}{[2 \pi f_\pi]^6} 
\int_{2 m_\pi}^{m_X} 
\left| 1+ \frac{3 m_\rho^2}{D_\rho(s)} \right|^2
k_\gamma^3 q_\pi^3 d\sqrt{s}~~~,~~~X=\eta,~\eta^\prime~~~.
\label{eq28}
\ee

This integration can be done by Monte Carlo techniques and gives~:
\be
\begin{array}{lll}
A_\eta = 38.25 \pm 1.07 ~~,~~{\rm eV} & A_{\etp} = 42.16 \pm 3.00 ~{\rm keV}~~~.
\end{array}
\label{eq29}
\ee

For further concern, one should note that these integrals are not
affected by the value for $x$.
Using the parameter values given in Table \ref{T0}, Eqs. (\ref{eq19}) and (\ref{eq10}), 
one can {\it compute} the partial widths and get the results collected in Table 
\ref{T2}. We have performed the computation by switching on/off the contact term 
contribution\footnote{When switching off the contact term in Eq. (\ref{eq28}), the numbers
in Eq. (\ref{eq29}) become $A_\eta =57.51 \pm 4.01$  eV and 
$A_{\etp}= 49.60 \pm 2.98$ keV, which is already conclusive.}. 
We stress again that all results presented in this Section do not depend on any free 
parameter and thus are {\it predictions} relying on the rest of the
HLS phenomenology.

\vspace{0.3cm}

\begin{table}[htb]
\begin{center}
\begin{tabular}{lrrr}
decay & PDG 2002 & Prediction with  & Prediction without \\ 
~~& ~~ & Box Anomaly & box Anomaly\\
\hline
$\eta \ra \pi^+ \pi^- \gamma$  (eV) & $55 \pm 5$  &  $56.3 \pm 1.7$ & $100.9 \pm 2.8$\\ [0.3cm]
Significance ($n ~\sigma)$ & ~~ &  $0.25 ~\sigma$  &  $8 ~\sigma$\\ [0.3cm]
\hline
$\eta^\prime \ra \pi^+ \pi^- \gamma$ (keV) & $60 \pm 5$  & $48.9 \pm 3.9 $ &$57.5\pm 4.0$\\ [0.3cm]
Significance ($n ~\sigma)$ & ~~ &  $1.75  ~\sigma$ & $0.39 ~ \sigma$\\ [0.3cm]
\hline
\end{tabular}
\caption{$\eta/\eta^\prime$ Partial widths as predicted by the HLS Model
when switching on/off the box anomaly contribution.
Significance is computed using an error obtained by adding
in quadrature the experimental error and the relevant model error
computed by Monte Carlo sampling (using information in Table \ref{T0}).}
\label{T2}
\end{center}
\end{table}

>From Table \ref{T2}, one clearly sees that the {\it predicted} partial width
for $\eta^\prime$ is not really sensitive to the presence of the  contact term. 
This can be well understood as, indeed, the value for $A_{\eta^\prime}$
is sharply dominated by the $\rho$ peak contribution provided by the $VVP$ 
Lagrangian term and the magnitude of the contact term is comparatively small.

In contrast, the {\it predicted} partial width for $\eta$ is much more
sensitive to the contact term because this contribution has only to
compete with the low mass tail of the $\rho$ distribution~; the
bulk of the resonance contribution is indeed sharply suppressed 
because the available phase space is small and located far outside
the $\rho$ peak.

Therefore, one can already conclude from  Table \ref{T2} that the
$\eta/\eta^\prime$ partial widths values provide a strong
evidence in favor of the box anomaly. Unexpectly, this evidence
is provided by the $\eta$ partial width alone. Additionally, the values
{\it predicted} for the box anomaly constants
$c_\eta \simeq 1.21 $  and $c_{\eta^\prime} \simeq 1.07$
from the rest of the HLS phenomenology fits nicely the $\eta/\eta^\prime$
partial widths, which means, for instance, consistency with having
$\theta_P = 10.30^\circ \pm 0.20^\circ$. 

Together with the results predicted for two--photon decay widths of 
the $\eta/\eta^\prime$ system, this also gives a strong support to
the  extended BKY breaking scheme summarized in Section \ref{breaking}
and to Eqs. (\ref{eq11}) and (\ref{eq21}) for the amplitude expressions at
the chiral limit.

\subsection{Invariant--Mass Spectra With/Without The Contact Term}
\label{shift}

\begin{figure}[htb]
  \centering{\
     \epsfig{angle=0,figure=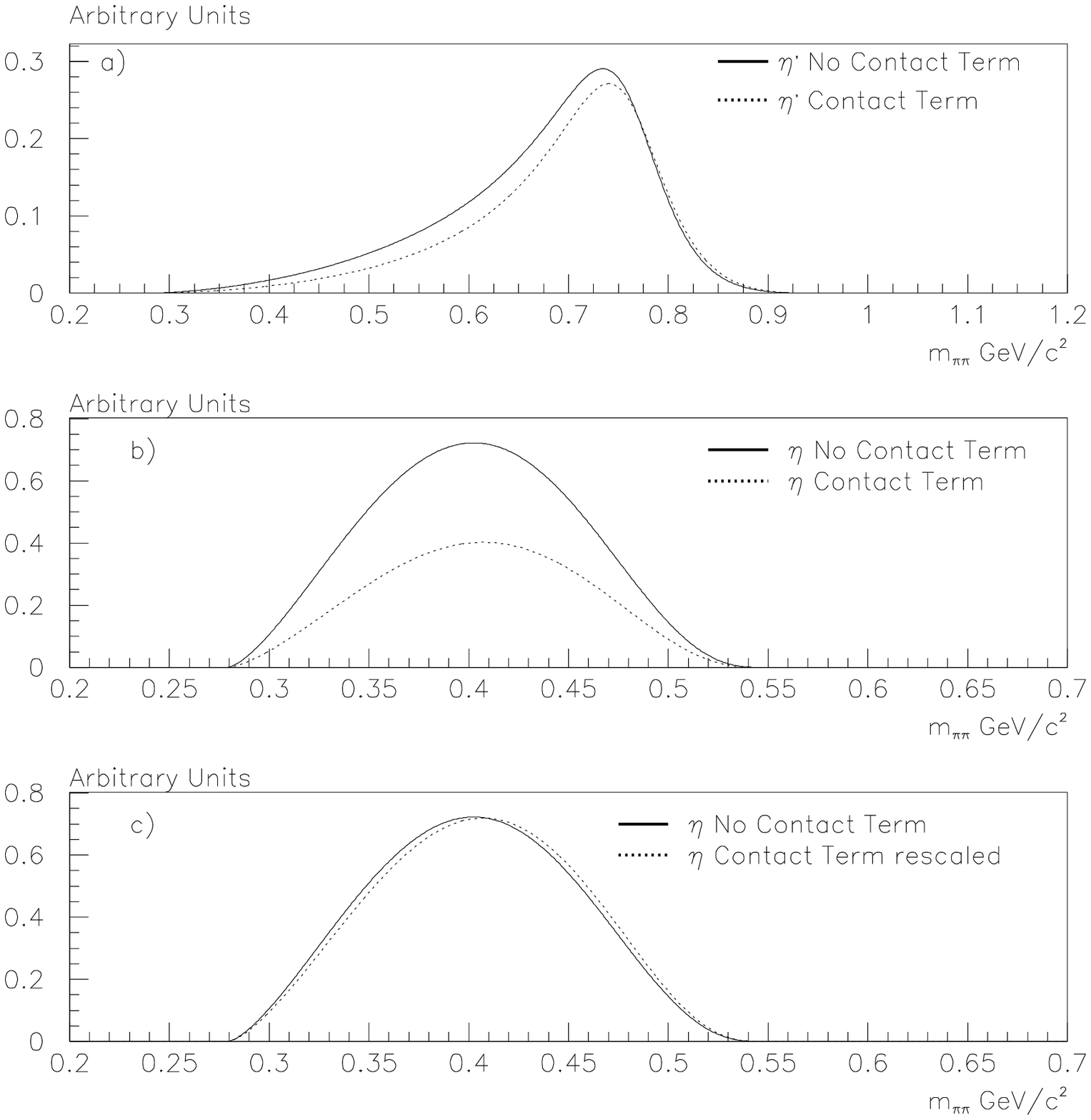,width=0.9\linewidth}
                    }
\parbox{150mm}{\caption ~~~
Predicted shapes for $\eta^\prime$ (top) and $\eta$ (mid)
distributions as functions of the dipion invariant mass.
Full line histograms correspond to having the contact term
in the amplitude, dotted line histograms correspond
to removing the contact term from the amplitude. 
 All other numerical parameters are at the same
values (see Table \ref{T0}). In the bottom figure, we
plot the prediction when accounting for the contact term (rescaled) 
superimposed with the prediction derived by removing this contribution.
\label{fig1}
}
\end{figure}

\indent \indent 
The shape of the dipion invariant--mass distributions are given in Figure \ref{fig1},
top for $\eta^\prime$, mid for $\eta$. These are proportional to yields
(up to acceptance/efficiency effects). The distributions are displayed
with having switched on/off the  contact term~; in these
two figures, the relative magnitude of the twin distributions is respected.

Looking at the $\eta$ distributions, one clearly understands the
width results given in Table \ref{T2}, as the integrals corresponding  
to box anomaly on/off are clearly very different (actually by a factor 
of about 2).

In the case of the $\eta$ meson, lineshape differences between
the case when the contact term is activated and
when it is dropped out are tiny as illustrated by Fig. \ref{fig1}, 
bottom. In this figure, one displays the distribution obtained
by removing the contact term and the one derived by activating it,
after rescaling it by $\simeq 1.8$.
 
The lineshape, in the case of the $\eta^\prime$, shows that
the peak location when accounting for the contact term is
slightly higher mass ($6 \div 8$ MeV) compared to the case when this
(CT) contribution is cancelled out. However, the main effect is that yields
below the $\rho$ peak location are somewhat suppressed because of the 
contact term.

In this sort of situation, if one performs a fit of an $\etp$ spectrum
affected by CT with only a resonance contribution and
lets free the resonance parameters, the shape will be distorted.
Indeed, in order to average reasonably the rising wing of the $\rho$ 
distribution, the peak has to be shifted to higher mass and therefore 
the observed mass must be larger. This is a mechanical effect connected
with the minimization of a $\chi^2$ for any appropriate function of one 
variable. We come back to this point when comparing with experimental data.

\section{ Experimental Data on $\eta/\eta^\prime \ra \pi^+ \pi^- \gamma$ Decays}
\label{data}

\indent \indent 
There are several sets of data available for the dipion mass
spectrum of the $\eta^\prime$ meson. Most of them  have been
published only as figures
\cite{Grigorian,TASSO,ARGUS,TPC2g,MARKIIa,LeptonF}.
For these, however, it happens that the information given in the 
body of the articles provides enough information in order to recover the 
yields and derive the acceptance/efficiency function~; the redundency
of the information is fortunately such that consistency checks and 
cross checks can be performed which validate the outcome of the
procedure. This is described in details in Section 4 of \cite{Ben0}
together with the peculiarities of each of these data sets.

Other spectra were available directly to us \cite{WA76} or as PhD 
theses \cite{Feindt,MARKIIc,McLean,Peters} published only as preprints~;
Here also the relevant information was either directly available or
could be reconstructed accurately, as for the references quoted above.
One should note that the data of \cite{McLean} supersed the 
ARGUS results published in \cite{ARGUS}.
 
These former data samples carry widely spread  statistics~;
474 events for the oldest data set \cite{Grigorian}, 130
events from TASSO \cite{TASSO}, 795 events from ARGUS
\cite{ARGUS} updated three years later to 2626 \cite{McLean},
321 events in the TPC--$\ggam$  sample \cite{TPC2g},
195 events in the PLUTO data set \cite{Feindt}, 586
in the CELLO data set \cite{Peters}, 401 for the data
set of WA76 collected using the Omega Prime Spectrometer 
at the CERN SPS \cite{WA76} and, finally, 2491 
(after acceptance corrections) for the experiment
performed at Serpukhov using the Lepton F facility \cite{LeptonF}.

The method used to extract the dipion invariant--mass spectra
from data is of special concern. These were derived from the
data samples just listed in the following way~: for each bin of 
dipion invariant--mass, one plots the $\pi^+ \pi^- \gamma$ 
invariant--mass spectrum and fits
with a gaussian (plus a polynomial background) the
number of $\eta^\prime$ it contains. In this way, one get
rid to a large extent of the precise background\footnote{
We have to make assumptions on the background shape across some 
small $\pi^+ \pi^- \gamma$ mass interval while the signal
is a narrow gaussian (typically 20 to 30 
MeV for its  standard deviation). This is certainly
much safer than assumptions on the background shape
over a 1 GeV invariant--mass interval with on top
of this  a signal as broad as a $\rho$ distribution.
} parametrization, as the signal is a narrow gaussian
peak dominated by the experimental resolution. 

 Performing this way, 
spectra appear without any background and the influence
of this in the data sets only reflects in the magnitude
of the errors on the yields per bin. It should be stressed
that this extraction method is obviously independent
of any assumption on the lineshape of
the underlying $\rho$ invariant--mass distribution.

For the data samples of \cite{LeptonF,WA76,Feindt}, the  
acceptance/efficiency function was directly known,
but without information on its uncertainties. As the 
spectra of \cite{WA76,Feindt}
carry small statistics, statistical errors are
dominant and errors on the acceptance/efficiency 
function can be neglected. For the data sample of
\cite{LeptonF}, the acceptance/efficiency function is 
provided as a curve (see their Fig. 3)~; as no information
is reported about uncertainties affecting this function,
these cannot be accounted for when folding  in
this function with any model distribution.

For the other data samples reviewed above, uncertainties 
on the acceptance/efficiency functions are also unknown,
as these can only be derived by unfolding it from the
fitting distribution. This was always provided as a product
of a well defined model function for the decay with
this acceptance/efficiency function. This is also
of little importance for all data sets dominated by
statistiscal errors, but it also affects the large
statistic sample of ARGUS we shall examine \cite{McLean}. 

Neglecting this source of uncertainties when computing
model errors certainly biases $\chi^2$ estimates towards
larger values (and smaller probabilities). However,
it should not spoil qualitatively model descriptions.

The sample of MarkII \cite{MARKIIa} is also significant
($\simeq$ 1200 events), however, the mass spectrum derived from this
has been obtained in a different way~: Selecting the events
in some mass interval around the $\eta^\prime$ mass in
the {\it global} $\pi^+ \pi^- \gamma$ invariant--mass spectrum, the corresponding
events are plotted in bins of $\pi^+ \pi^-$ invariant--mass. 
This spectrum is then described as a superposition of a
$\rho$ mass distribution plus some background, and a global fit
to this spectrum provides the signal ($\rho$)  and background
populations inside each bin. Therefore this method assumes 
an accurate knowledge of all phenomena contributing to the
background (and of its parametrization)~; it also relies 
on the way the $\rho$ lineshape is  parametrized.
This is also, basically, the method used to study the $\eta^\prime$ 
mass distribution performed by the L3 Collaboration \cite{L3}
on a sample of $2123 \pm 53$ events~; this will be specifically 
discussed at the appropriate place below, as it is the latest 
published data sample.

Some other papers published spectra without background subtraction
(namely\cite{MARKIIb} and \cite{MARKIIc} which carry actually the
same data). In order to use these, one would have to model the
background without any motivated knowledge of the data 
set and detector properties\footnote{Indeed, beside extracting
the yields, one needs to estimate the acceptance/efficiency
function which might well be different for signal and
background events.}~; therefore, this MarkII spectrum 
will not be examined here. This lack of background subtraction 
is also the reason why the spectrum published by the L3 
Collaboration is also skept.

Finally, the most reliable spectrum for the $\eta^\prime$
decay is the one collected by the Crystal Barrel Collaboration
\cite{Abele} which is also, by far, the largest data sample
(7392 events). This spectrum has been constructed using
the method described at the beginning of this Section~;
therefore, it is independent  of any assumption on the 
underlying $\rho$ invariant--mass distribution and, thus, 
is certainly free from any prejudice or bias. Additionally, 
this data sample is certainly the most secure to be used
as uncertainties on acceptance and efficiencies are already 
included into yield errors and, thus, model comparison can be 
performed directly and reliably.

\vspace{0.5cm}

The corresponding spectra for the $\eta$  decay,
have been derived from $\pi p \ra \eta n$ data
collected long ago by two experiments \cite{Gormley,Layter}~;
the published data are already background subtracted.
They both carry large statistics (7250 for \cite{Gormley} 
and 18150 for \cite{Layter}). The relevant results have been 
published only as figures. Yields per bin can be read off from 
these figures without any difficulty together with their errors.

In \cite{Gormley}, the acceptance/efficiency function
is provided directly and also folded in with a well defined
model function. It is given superimposed to the case when
the analysis is performed with a simple gauge--invariant
phase--space matrix element and with a $\rho$ dominant
one. The two corresponding acceptance/efficiency functions
are conflicting, essentially in the region 
$k_\gamma = 90 \div 110$ MeV. However, this seems
to reflect their dependence upon angular distributions.
It is therefore worth using the functional information
associated with the $\rho$ dominant matrix element.

For the data of Layter {\it et al.} \cite{Layter}, the
acceptance/efficiency function is not shown and should
be unfolded from the theoretical $\rho$ (and phase--space) 
distribution(s). This information can be extracted 
with some reliability~; in contrast with  \cite{Gormley}, 
this  yields a function extracted from the $\rho$ distribution 
very close from those extracted from the simple phase space 
distribution. Actually, this data set should be considered with 
some care as extracting the acceptance/efficiency function
can only be performed by making some assumption on the $\rho$
mass actually used in this paper \cite{Layter}. We have conservatively
assumed that Layter {\it et al.} \cite{Layter} used the same $\rho$ mass 
as Gormley {\it et al.} \cite{Gormley}, namely $m_\rho= 765 $ MeV~;
this assumption is crucial and cannot be ascertained. This
makes more secure information derived from the Gormley 
spectrum. 

Finally, for both $\eta$ spectra, it is impossible to restore
the accuracy on the acceptance/efficiency function.
These will be considered negligible in the present 
study.

\section{ Experimental Data Versus Predictions For $\eta/\eta^\prime$ Spectra}
\label{shapes}

\indent \indent 
As seen in Subsection \ref{function}, the HLS Model provides 
definite spectra for 
both $\eta/\eta^\prime$ invariant--mass distributions. These 
depend on parameters which can be fixed independently of
the $\eta/\eta^\prime \ra \pi^+ \pi^- \gamma$  spectra,
like $a$, $g$, $x$ (see Subsection \ref{numbers}), and of
the $\rho$ propagator which is fitted elsewhere \cite{ff1}
with parameters values as determined in these fits (see 
Eq. (\ref{eq26}) and Table \ref{T0}). The model fairly predicts  
the absolute magnitude (the integral) of each spectrum as illustrated in 
Table \ref{T2}.
In this Section, we   
focus on comparing the {\it predicted} lineshapes derived from 
the model Eq.(\ref{eq24}) with  the data listed in Section \ref{data}. 

As all data considered are binned, we have integrated the predicted
function Eq. (\ref{eq24}) (or Eq. (\ref{eq28})) over the bin size
and normalized this function to the integral of the experimental
distribution. When relevant ({\it i.e.} all spectra except for the one
of Crystal Barrel \cite{Abele}), the model function was folded in
with the acceptance/efficiency function derived for each 
of the above data samples.

\begin{figure}[htb]
  \centering{\
     \epsfig{angle=0,figure=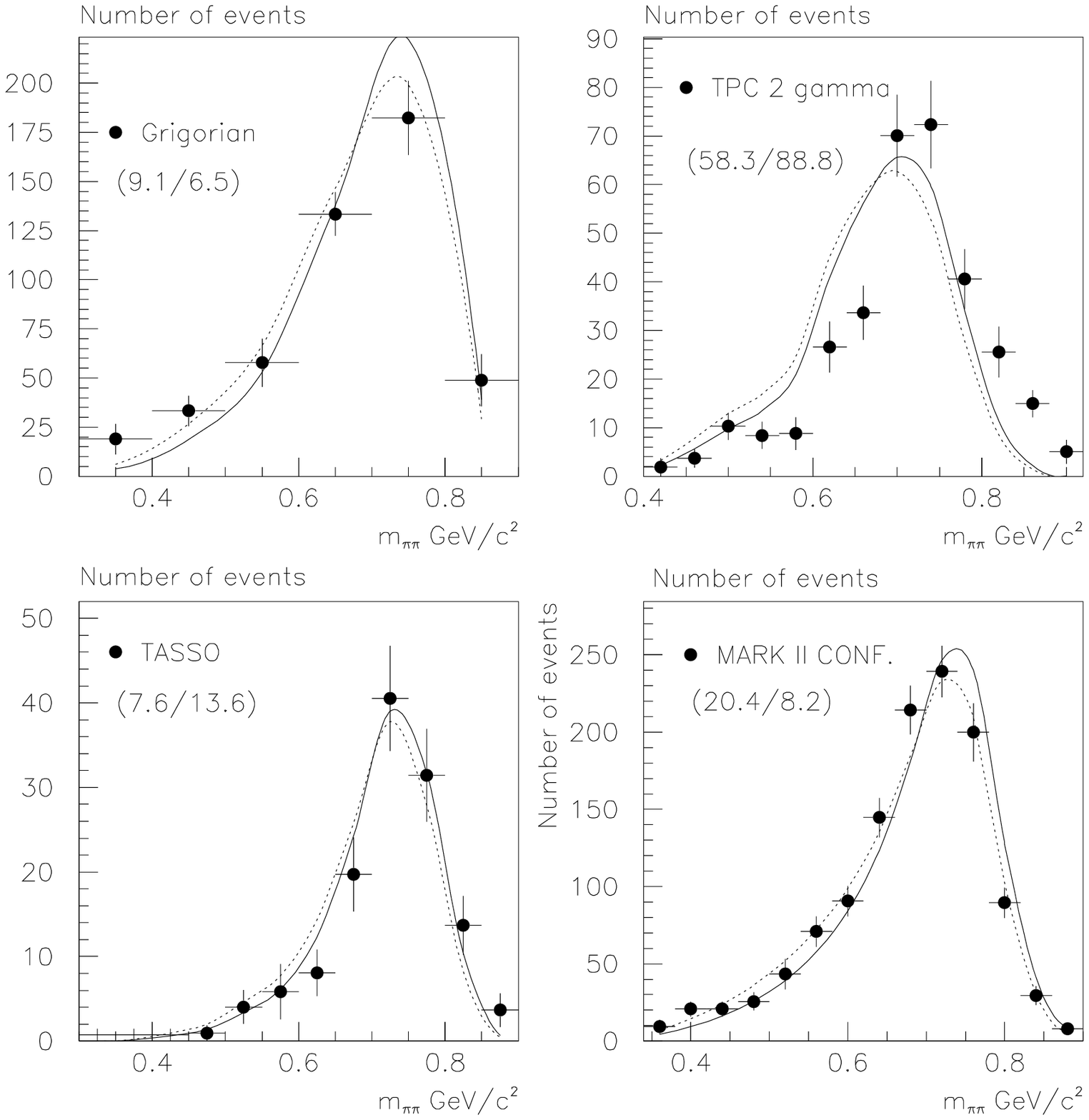,width=0.9\linewidth}
                    }
\parbox{150mm}{\caption ~~~
Invariant dipion mass Distributions for $\eta^\prime$ decay.
Experimental data sets with the predicted distributions
without the contact term (dashed curve) and with
this contribution activated (full curve). The numbers given
are $\chi^2({\rm contact~term})/\chi^2({\rm no~contact~term })$
for the lineshapes only.
\label{fig2}
}
\end{figure}

\begin{figure}[htb]
  \centering{\
     \epsfig{angle=0,figure=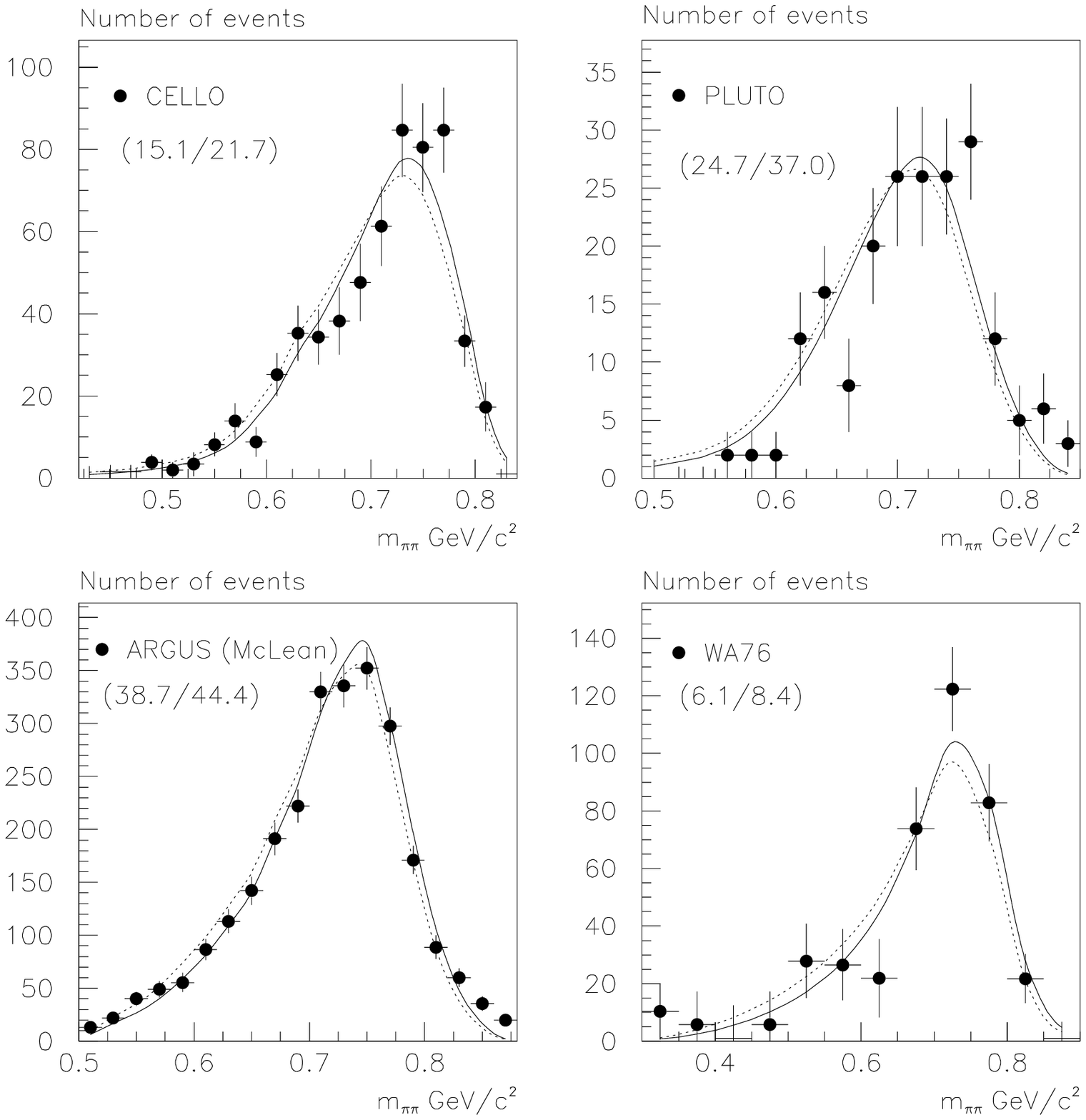,width=0.9\linewidth}
                    }
\parbox{150mm}{\caption ~~~
Invariant dipion mass Distributions for $\eta^\prime$ decay.
Experimental data sets with the predicted distributions
without  the contact term  (dashed curve) and with
 this contribution activated (full curve). The numbers given
are $\chi^2({\rm contact~term})/\chi^2({\rm no~contact~term })$
for the lineshapes only.

\label{fig3}
}
\end{figure}

\begin{figure}[htb]
  \centering{\
     \epsfig{angle=0,figure=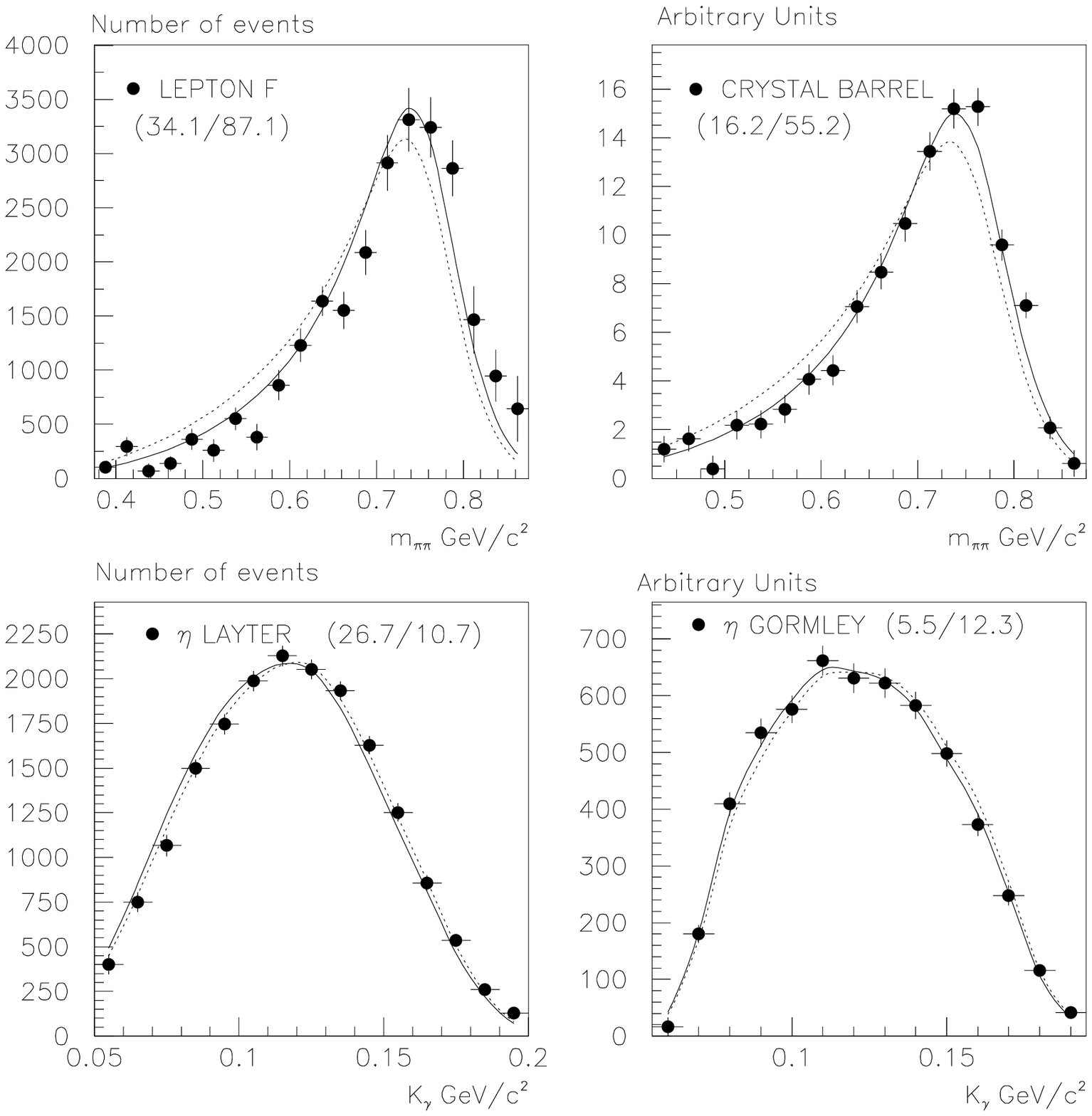,width=0.9\linewidth}
                    }
\parbox{150mm}{\caption ~~~
Invariant dipion mass Distributions for $\eta^\prime$ decay
and the single $\eta$ decay (as a function of the photon momentum
in the $\eta$ rest frame). Experimental data sets with the predicted 
distributions without  the contact term  (dashed curve) and with
 this contribution activated (full curve). 
The numbers given
are $\chi^2({\rm contact~term})/\chi^2({\rm no~contact~term })$
for the lineshapes only.
\label{fig4}
}
\end{figure}

The results are displayed in Figs. (\ref{fig2}) to (\ref{fig4}).
One should note that all $\eta^\prime$ spectra are given as functions
of the dipion invariant--mass, while $\eta$
spectra are given as functions of the photon momentum in the 
$\eta$ rest frame.

In these figures, together with the specific experimental
spectrum, we show the predicted curve (computed just as defined
just above) when keeping the contact term (full
curve) and the one derived by dropping out this contribution
(dashed curve). These two cases will be referred to as 
resp. {\bf CT} and {\bf NCT}. In these figures, we give
the $\chi^2$ corresponding to these two solutions under the 
form $\chi^2({\rm CT})/\chi^2({\rm NCT})$.
The number of degrees of freedom can be easily read off
from the spectra as this is exactly the number of bins
of the experimental histogram.

The curves shown have been computed at the central
values of the parameters as given in Table \ref{T0}.
The $\chi^2$'s have been computed by folding in the experimental
error in each bin with a model error also computed bin per bin.
These model errors have been computed by sampling the parameters
around their central values with standard deviations given
by their quoted errors (see Table \ref{T0}). Except for
Crystal Barrel \cite{Abele}, where it is irrelevant,
uncertainties on the acceptance/efficiency functions
are not (cannot be) accounted for. The curves
shown are actually histograms which have been smoothed
automatically by the {\sc hbook/paw} package.

\vspace{0.5cm}

Examination of  Figs. (\ref{fig2}) to (\ref{fig4}) is quite
interesting. First of all, the spectrum from TPC--$\ggam$
is clearly the single one far away from predictions, indicating 
that something was not well controlled when extracting it from data. 
All others match well, or quite well, the predictions~; this 
clearly gives support to the model developped in the previous
Sections and to the relevance of the parameters  given
in Table \ref{T0}.

In terms of probabilities (reflected by the $\chi^2/dof$ values 
given in  the figures), the oldest data set of \cite{Grigorian} gives 
comparable probabilities to either of the 
{\bf CT/NCT} assumptions, while maybe slightly prefering
the {\bf NCT} assumption. {\bf CT/NCT} descriptions are
practically equivalent for the ARGUS
\cite{McLean} and WA76 \cite{WA76} spectra, while
nevertheless slightly favoring the {\bf CT} assumption.

The relatively low statistics spectra provided by 
TASSO \cite{TASSO} ($\chi^2$ ratio of 0.6 in favor
of the {\bf CT} assumption), CELLO 
\cite{Peters} (0.7) and PLUTO \cite{Feindt} (0.7) 
somewhat prefer the {\bf CT} assumption.

Finally,
the two largest statistics experiments Lepton F
\cite{LeptonF} and Crystal Barrel \cite{Abele} sharply
favor both the {\bf CT} assumption
against {\bf NCT}~; the $\chi^2$ 
distance is indeed better by more than a factor
of 2 .

To be more precise the Lepton F spectrum\footnote{
The 8 lowest mass points of this spectrum
contribute severely to the $\chi^2$ for
both ({\bf CT/NCT}) assumptions. On the other hand, the 
sharp drop in acceptance \cite{LeptonF} at large 
$m_{\pi \pi}$ might have been difficult
to estimate reliably. Qualitatively, however, the clear 
preference of this distribution for the {\bf CT} assumption
is obvious.}
gives a 3\% probability to the {\bf CT} assumption
and a $2~10^{-8}$ \% probability to the {\bf NCT} assumption.
These relatively low probabilities should be related
with the lack of information on the acceptance/efficiency
function which affects in a same manner both solutions.
Accounting for the corresponding errors would certainly 
increase both probabilities but hardly switch 
their ordering.

The corresponding probabilities for the Crystal
Barrel data set are resp. 57.8\% and 0.1 \%~; 
these values are certainly realistic as the model
errors are reasonably well accounted for.
 
Among these two data sets which could be used as reference,
the Crystal Barrel data (available in a directly usable
form \cite{Abele}) should clearly be prefered, as systematics
are better controlled all along the invariant--mass
 range. It also carries, by far, the largest statistics.

\vspace{0.5cm}

>From $\chi^2$ values,  the shape of the $\eta$ 
spectrum from Layter {\it et al.} \cite{Layter} 
seems in better agreement with the {\bf NCT} assumption
(77\% probability) than with the {\bf CT} assumption 
(3\% probability). 

In contrast, the description of the $\eta$ spectrum
from Gormley {\it et al.} \cite{Gormley} is simply perfect 
and corresponds
to resp. 97.7 \% probability for the {\bf CT} assumption
and to 58.2 \% probability for the {\bf NCT} assumption~;
this reflects better the remark following from Fig. \ref{fig1}
(bottom) that these lineshapes are very close together.
The $\chi^2$ values, however, indicate that the
geometrical distance ($\simeq \sqrt{\chi^2}$)
of this spectrum to the {\bf CT} solution
is significantly smaller than those to
{\bf NCT}.
 
\begin{figure}[htb]
  \centering{\
     \epsfig{angle=0,figure=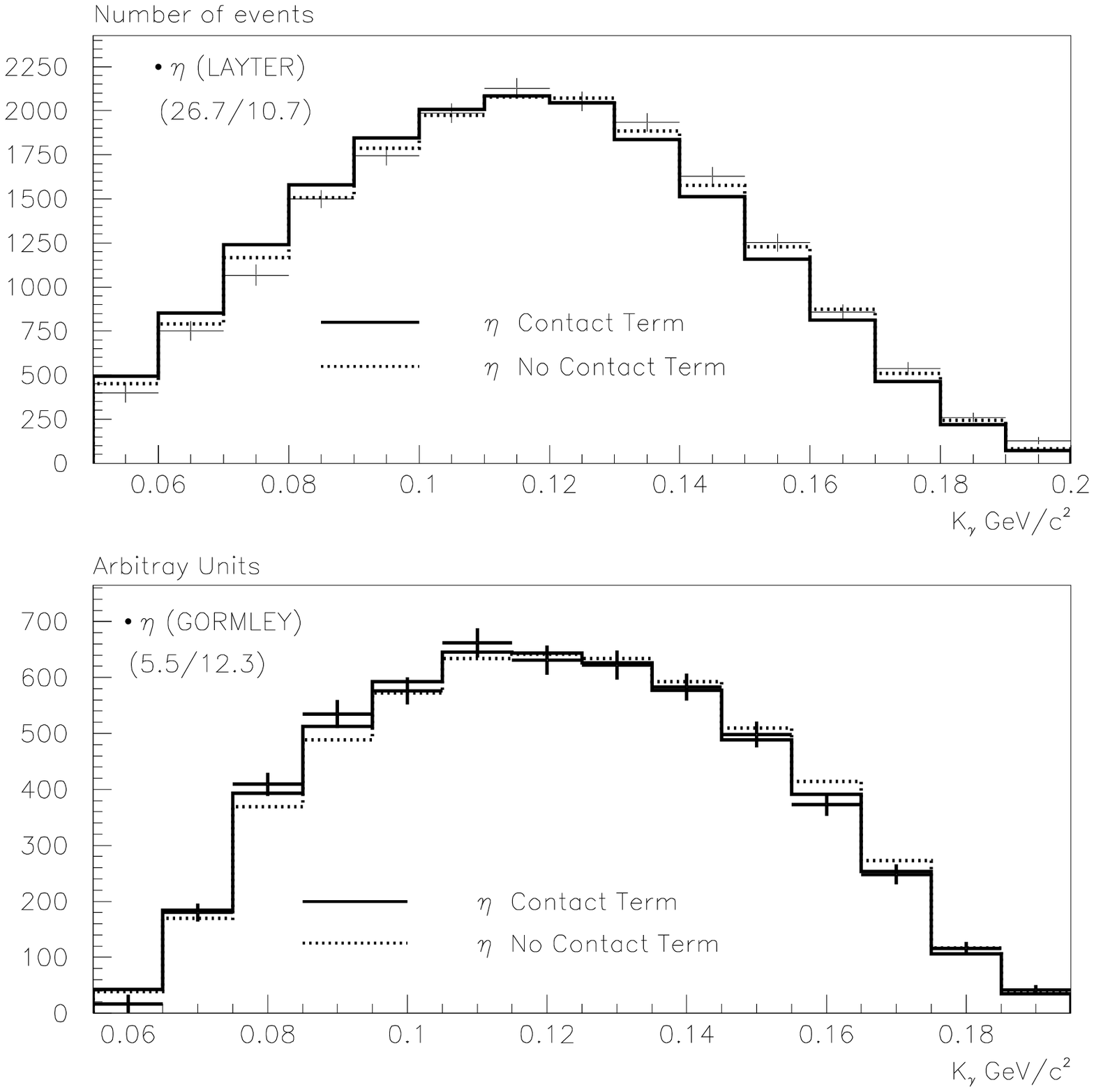,width=0.9\linewidth}
                    }
\parbox{150mm}{\caption ~~~
Photon momentum distribution in $\eta$ decay
Experimental data are from Layer {\it et al.} \cite{Layter}
(top), and Gormley {\it et al.} \cite{Gormley} (bottom)~;
Experimental data sets with the predicted distributions
without  the contact term  (dashed curve) and with
 this contribution activated (full curve). 
The numbers given
are $\chi^2({\rm contact~term})/\chi^2({\rm no~contact~term })$
for the lineshapes only.
\label{fig5}
}
\end{figure}

Fig. \ref{fig5} gives the same information as in
Fig. (\ref{fig4}) but enlarged and binned. 
Here one sees that a third of the $\chi^2$ 
for the Layter spectrum \cite{Layter} comes from only
the bin covering the momentum interval $60 \div 80$ MeV/c.
Compared to the same result for the Gormley spectrum
\cite{Gormley}, the Layter spectrum looks a little bit skewed.
It is, however, impossible to decide whether this comes
from systematics affecting the acceptence/efficiency function
as this (skewed) shape happens to match nicely  the
{\bf NCT} assumption\footnote{
This skewness might have been magnified unwillingly
by  the choice of the $m_\rho$  value  we performed in
order to  extract the acceptance efficiency/function for 
the Layter spectrum. Any underestimation of this
input $m_\rho$ value contributes to the  skewness of 
this distribution.
We are responsible for this uncertainty, but we did not
find an unbiased way out.}.

However, the $\eta \ra \pi^+ \pi^- \gamma$  partial width alone 
\cite{PDG02}, certainly a more secure information,
and the Gormley spectrum   
undoubtfully favor the {\bf CT} assumption against the 
{\bf NCT} one. These two aspects have to be balanced in a global 
fit accounting for lineshapes and partial widths. 

\section{A Global Fit to $\eta/\eta^\prime$ Spectra and Widths}
\label{globalfit}

\indent \indent
 We have compared the data (lineshapes and partial
 widths) with the predictions of our model  fed with
 numerical and functional information coming from the rest
 of the phenomenology accessible to the HLS framework,
 without any parameter freedom. The results obtained 
 in Section \ref{prediction} and \ref{shapes} considered
 together indicate that the model is valid and favors the 
 the contact term  as a physically motivated
 contribution to decay processes. We remind that this contact term
 is $not$ a free parameter, as widely discussed above.
 
In view of this, it looks worth performing a simultaneous fit
of the $\eta$ data sets with some accurate $\eta^\prime$ 
spectrum~; for reasons explained above,  it is certainly worth choosing 
the Crystal Barrel spectrum.  As a clear conclusion should take 
into account all aspects of the available experimental information, 
partial widths have been fed into the $\chi^2$ to be minimized.

\vspace{0.3cm}

In order to 
perform this fit one needs to release some of the parameters
fixed as in Table \ref{T0}~; as the main information
for the present purpose is the peak location, it looks 
worth releasing the parameters named $e_1$ and $e_2$ which mostly
influence the $\rho$ peak location. Comparing the values
returned from this fit to the corresponding values originally 
extracted from fitting the pion form factor could contribute
to clarify the conclusion, as one can consider the $\rho$
propagator as a universal function, as valid for $F_\pi(s)$
as for the the $\eta/\etp$ spectra.

Indeed, we know that, in the pion form factor, the subtraction
polynomials of $\Pi_{\rho \rho}(s)$ and $\Pi_{\rho\gamma}(s)$
are somewhat competing and that some (small) correlation
among the corresponding polynomials exists \cite{ff1}~; it is therefore
motivated to attempt freeing $e_1$ and $e_2$ as these
correlations could have spoiled their central values 
by some (certainly) small amount. However, the parameter
values returned from fit must not be inconsistent with
their partners derived from fit to the pion form factor only\footnote{
This actually means that a further test could be a simultaneous
fit, within a consistent framework, of the pion form factor
and of the relevant $\eta/\etp$ decay information. One 
does not expect neither a surprise nor hard difficulties from such
an attempt~; the present work  indicates that this should not 
provide more insight than a global probability.
}.

On the other hand, the other parameter values in Table
\ref{T0} describing fairly well the full set of $V \ra P \gamma$
and $P \ra V \gamma$ decays  would hardly accomodate
a significant change of their values without failing to fit
the $VP\gamma$ processes.

In order to avoid too much correlations which can hide clarity
in the conclusions, we shall test separately the {\bf CT} 
and the {\bf NCT} assumptions. Indeed, as the HLS Model
predicts the magnitude of the constant contact term (if any), 
it seems enough to check its precise relevance and no attempt
will be made to fit its value. Finally, for the present exercise,
we neglect model errors\footnote{ These are certainly 
present as the uncertainties on $a$, $g$ and $x$ 
contribute to model errors, even when releasing any 
constraint on $e_1$ and $e_2$.}~; this mechanically
makes the $\chi^2$'s slightly more pessimistic than 
they really are.

\vspace{0.3cm}

We could have chosen to perform a simultaneous fit
of the Crystal Barrel $\etp$ data set \cite{Abele} together with both
$\eta$ data sets \cite{Layter,Gormley} simultaneously. 
One could indeed imagine that the systematics could compensate. 
We have, nevertheless,  prefered performing 
the fits separately for the Crystal Barrel $\etp$ spectrum 
together with each of these $\eta$  spectra in isolation. 
Using both $\eta$ spectra certainly leads to  intermediate 
fit qualities.

Additionally, before letting $e_1$ and $e_2$ vary, we have 
performed the ``0 parameter fit'' in order to get the $\chi^2$'s
and probabilities when using directly the parameter values 
as given in Table \ref{T0}. In this way, we 
know the starting quality of the global description of these
decay modes induced by the rest of the HLS
phenomenology~; we can also estimate what is gained by letting
some parameters to vary. 

\vspace{0.3cm}

When using the data set of Layter {\it et al.} \cite{Layter},
while accounting for the $\eta/\etp$ contact terms  at the expected level
(assumption {\bf CT}), one clearly sees from Table \ref{T3}
that the $\rho$ lineshape parameters $e_1$ and $e_2$ 
do not move farther than 2 $\sigma$ from the values
found when fitting the pion form factor \cite{ff1}
(the present  $\sigma$'s are, however, much larger than 
found in fits to $F_\pi(s)$ \cite{ff1}).
The gain in $\chi^2$ got by releasing these parameters
is modest (2.7) and the central values for the partial widths
get a little changed. This confirms that
the parameter values in Table \ref{T0} giving
$\chi^2/dof=37.88/35$ (34\% probability) are already close 
to optimum~; leaving them free, essentially improves
the $\eta$ spectrum lineshape slightly, but at the 
expense of slightly degrading the central values
of the partial widths.

\begin{table}[htb]
\begin{center}
\begin{tabular}{lccccc}

Layter \cite{Layter} &   ~~      &    ~~  &   ~~    & ~~  & ~~   \\[0.3cm]
\hline
~~~~                &             $e_1$      &     $e_2$              &  $\chi^2/dof$  	& $\eta$    & $\etp$    \\
~~~~                &   ~~~                  &   (GeV$^{-2})$         &  (Prob.)      	& P.W. (eV) &P.W. (keV) \\[0.3cm]
\hline
CT $+$ No Fit      &    $~~0.222 \pm 0.011$ &     $-1.203 \pm 0.017$ &   $37.88/35$    &  56.3     &   48.9    \\
~~~~                &   ~~~                  &  ~~~                   &    (34\%)       & ~~~  &             \\[0.3cm]
CT $+$ Fit         &  $~~0.339^{+0.105}_{-0.056}$   &  $~-1.395^{+0.143}_{-0.160}$     &   $35.16/33$    &  53.3  & 46.2 \\
~~~~                &   ~~~                  &  ~~~                   &    (36.6\%)       & ~~~  &                 \\[0.3cm]
\hline
No CT $+$ No Fit   &    $~~0.222 \pm 0.011$ &     $-1.203 \pm 0.017$ &   $140.65/35$    &  100.9     &   57.5    \\
~~~~                &   ~~~                  &  ~~~                   &    (0 \%)        & ~~~  &             \\[0.3cm]

No CT $+$    Fit   & $~~0.933^{+0.321}_{-0.093}$ &   $-2.355^{+0.180}_{-0.200}$ &   $60.88/33$    &  75.3     &   40.0    \\
~~~~                &   ~~~                  &  ~~~                   &    (0.2 \%)        & ~~~  &         \\[0.3cm]
\hline
\hline
Gormley \cite{Gormley} &   ~~      &    ~~  &   ~~    & ~~  & ~~    \\[0.3cm]
\hline
~~~~                &  $e_1$    &     $e_2$       &  $\chi^2/dof$ & $\eta$    & $\etp$ \\
~~~~                &   ~~~     &   (GeV$^{-2})$  &  (Prob.)      & P.W. (eV) &P.W. (keV) \\[0.3cm]
\hline
CT $+$ No Fit      &    $~~0.222 \pm 0.011$ &     $-1.203 \pm 0.017$   &   $25.58/34$    &  56.3      &   48.9       \\
~~~~                &   ~~~                  &  ~~~                     &    (85\%)       & ~~~  &             \\[0.3cm]
CT $+$ Fit         &  $~~0.269 \pm 0.080$   &  $-1.275^{+0.135}_{-0.155}$     &   $25.01/32$    &  54.6     &   47.7 \\
~~~~                &   ~~~                  &  ~~~                   &    (80.6 \%)       & ~~~  &              \\[0.3cm]
\hline
No CT $+$ No Fit   &    $~~0.222 \pm 0.011$ &     $-1.203 \pm 0.017$   &   $54.13/34$    &  76.9      &   53.4       \\
~~~~                &   ~~~                  &  ~~~                     &    (1.6\%)       & ~~~  &            \\[0.3cm]

No CT $+$    Fit   &    $~~0.529 \pm0.090$ &     $-1.700^{+0.154}_{-0.195} $   &   $36.38/32$    &  65.7      &   44.5       \\
~~~~                &   ~~~                  &  ~~~                     &    (27.2\%)      & ~~~  &           \\[0.3cm]
\hline
\hline
\end{tabular}
\caption{
Simultaneous fits of the $\eta/\etp$ distributions from \cite{Abele,Layter}
on the one hand, and from \cite{Abele,Gormley} on the other hand.
CT stand for the ``contact terms'' generated by the box part
of the WZW Lagrangian (see Eq.(\ref{eq5})).
The values for $e_1$ and $e_2$
quoted in  ``No Fit'' entries are taken from Table \ref{T0}
and not varied from their central values. The `` P.W.'' entries
are the central values for the $\eta$ and $\etp$ partial widths
in the appropriate units~; the recommended values for these \cite{PDG02}
are given in Table \ref{T2}.}
\label{T3}
\end{center}
\end{table}

When dropping out the contact term contributions
(assumption {\bf NCT}),  the  $\rho$ lineshape parameters 
$e_1$ and $e_2$  change significantly with respect
to their starting point, with much larger errors
than originally. Even then, the fitted values
for $e_1$ and $e_2$ move by more than 6 (new) $\sigma$
from expectations~; therefore, these fitted values
can be considered inconsistent with their values fitted in the
pion form factor. Additionally, even if the gain is
large ($\chi^2/dof$  is improved from 140.65/35 
to 60.88/33), it is not sufficient to push
the probability (0.2\%)  to a reasonable value.
Therefore, the peculiar uncertainties affecting
this spectrum does not prevent to reach a clear global 
conclusion.

\vspace{0.5cm}

When using the data set of Gormley {\it et al.} \cite{Gormley}
together with the $\etp$ data of Crystal Barrel, the picture
is unchanged, but looks much clearer. Fixing the parameters
to their values in Table 1, gives already a remarkable fit
quality (probability 85 \%), when contact terms  are accounted
for. In this case,  letting free $e_1$ and $e_2$, the $\chi^2$  
improves a little (0.57 unit), but the fit probability
degrades to  80 \%, because of the smaller number
of degrees of freedom. The $\rho$ parameters move by about
1 (new) $\sigma$ from expectations and thus stay
consistent with the $\rho$ lineshape determined
when fitting the pion form factor \cite{ff1} (the
region of the minimum $\chi^2$ seems flat).

When removing the contact term from our expressions
({\bf NCT} assumption), the starting values of the fit 
parameters provide a comparatively poor decription 
(1.6 \% probability)
and the $\eta$ partial width is far from expectations
\cite{PDG02} (see also Table \ref{T3}). Now, releasing 
$e_1$ and $e_2$ improves significantly the description,
as we reach a 27 \% probability after fit, with
reasonable central values for both partial widths.  
The price to be paid for this configuration is that the 
parameters $e_1$ and $e_2$ change by more than 3 (new) 
$\sigma$ from  expectations. Therefore, the 
{\bf NCT} assumption returns a $\rho$ lineshape
inconsistent with fits to the pion form factor.

\vspace{0.3cm}

>From the previous Sections, we already knew that 
the {\bf CT} assumption is certainly favored in a global
account of both shape and partial width for both
the $\eta$ and $\etp$ meson simultaneously. We also knew
that the {\bf NCT} assumption was disfavored under
the same conditions. 

What we have learnt in this Section is that,
in order to accomodate the description
of all aspects of the $\eta/\etp$  information, 
the {\bf NCT} assumption gives up being consistent with 
the $\rho$ lineshape as found by fitting the pion form 
factor \cite{ff1}.

Therefore, we conclude that experimental
data do provide a fair evidence in favor of the
box anomaly phenomenon at the expected level~; additionally, the
sharing observed between resonant and contact
term contributions ($-3 : 1$) is well predicted by
the FKTUY assumption \cite{FKTUY} leading to
the Lagrangian in Eq. (\ref{eq5}). 

\vspace{0.3cm}

\begin{figure}[htb]
  \centering{\
     \epsfig{angle=0,figure=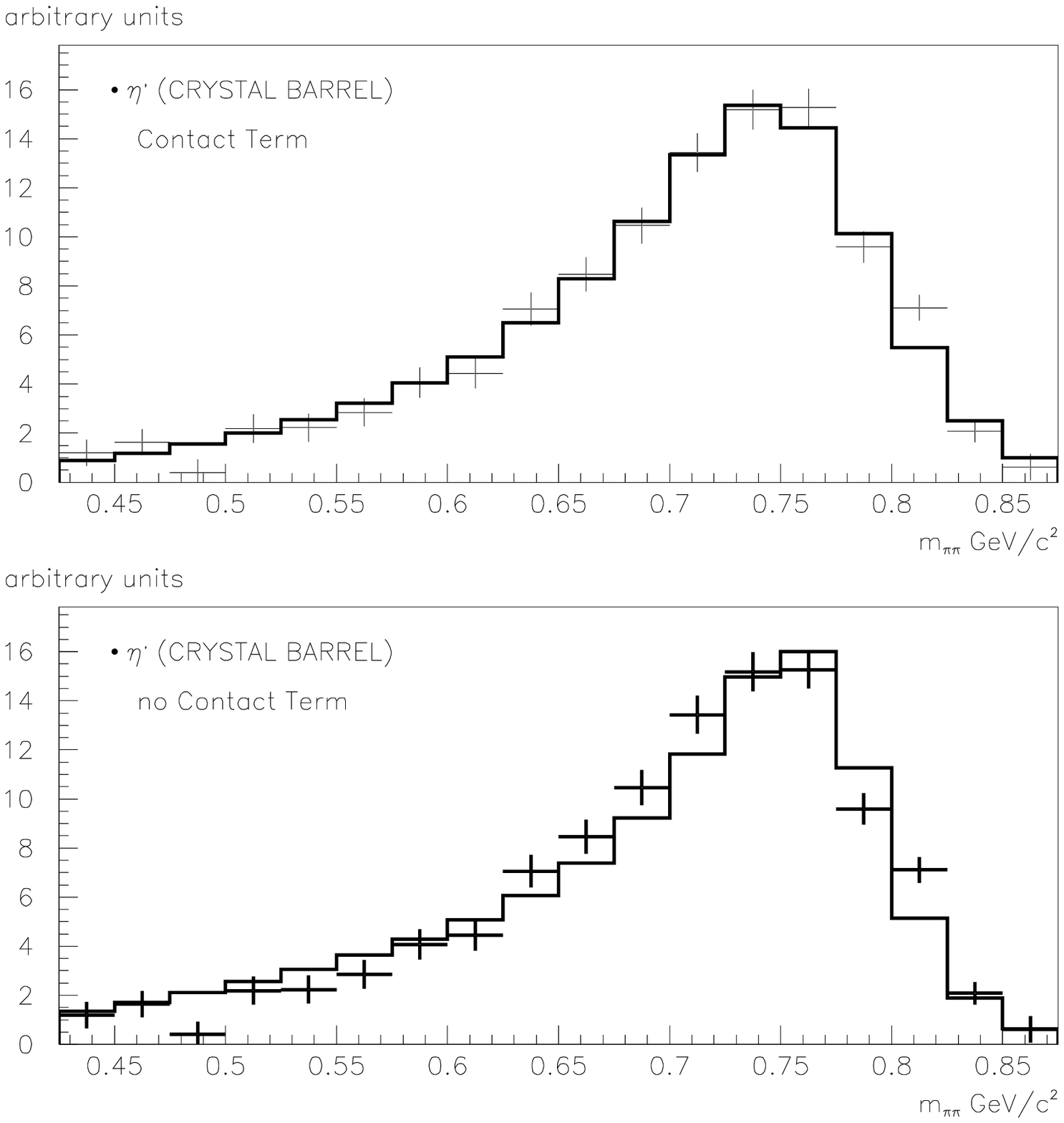,width=0.9\linewidth}
                    }
\parbox{150mm}{\caption ~~~
Fit of the $\etp$ invariant mass spectrum, top by including
the contact term, bottom by removing
this term. Compare the  peak locations.

\label{fig6}
}
\end{figure}

In Fig. (\ref{fig6}), we show  the description
of the Crystal Barrel spectrum \cite{Abele}
using  Eq. (\ref{eq24}) (or Eq. (\ref{eq28}))
with the contact term considered and
removed. In order to get this we performed
fits leaving free $e_1$ and $e_2$.

When accounting for the $\etp$  contact term, the $\rho$ 
peak location is found in the bin covering the mass region 
from 725 to 750 MeV. When dropping it out,  the (fit) mechanism 
described in Subsection \ref{shift} makes the $\rho$ peak 
shifting to the next bin which  covers the mass interval
from 750 to 775 MeV. This trend was already observed
by \cite{LeptonF,Abele} and also by most
Collaborations who have performed extraction
of the $\etp$ spectrum canonically~; sometime too 
much \cite{TPC2g}. 

A real shift exists and is small (see Section \ref{shift}).
It is artificially increased by the fit procedure
in order to get a better account of the low mass tail
of the $\etp$ invariant--mass spectrum. However, this
artificial large mass shift is indeed the signal of the
box anomaly.
\vspace{0.3cm}

It is claimed in \cite{L3} that the L3 Collaboration
does not observe a $\rho$ peak shift. Several reasons 
can be invoked. First, as remarked above, \cite{L3} did not 
perform the $\etp$  spectrum extraction canonically and, therefore,
any conclusion about the underlying $\rho$ lineshape 
in the $\eta^\prime$ decay becomes a delicate matter.

Among other reasons the most likely is  their fitting of 
the $\rho$ mass and width.  The $\rho$ (Breit--Wigner) mass 
is thus found at values ($766 \pm 2 $ MeV) normally obtained
in only processes where the dynamics is not really
well under control (hadroproduction or 
photoproduction) \cite{PDG02}. 

If, for a moment, the L3 result were considered as reference
in order to detect a $\rho$ peak shift, 
one might have instead to consider that a shift occurs in
$e^+e^-$ annihilations or $\tau$ decays, as  these yield   
rather larger $\rho$ (Breit--Wigner)  masses ($ \simeq 775$ MeV) 
\cite{PDG02}. Under these conditions, 
it is difficult to draw any conclusion from \cite{L3} about 
the existence (or absence) of a $\rho$ peak location shift 
in the $\etp \ra \pi^+ \pi^- \gamma$ decay.

\section{Fits To The Four Anomaly Equations}
\label{anomalies}

\indent \indent
>From now on, we make the assumption that the correct
set of equations defining the anomalous amplitudes at the 
chiral limit are given by Eqs. (\ref{eq11})  and (\ref{eq21}) and 
no longer by Eqs. (\ref{eq3}). These have been derived
using the approximate field transformation Eq.(\ref{eq8}).
We also examine, for completeness, the case when
the exact field transformation is used~; the way to modify
our anomaly equations to go from one case to the other
is given in the Appendix.  

In both cases, these equations actually depend on
only one parameter (resp. $x$ or $\lambda$)~; this
can legitimately look like a severe constraint.

\vspace{0.3cm} 

\begin{table}[htb]
\begin{center}
\begin{tabular}{lrrc}
~ & PDG 2002  & Fit Result  & Significance (n $\sigma$) \\
\hline
$x$        		& ~  & $0.911 \pm 0.015 $ & ~~~\\ [0.25cm]
$\chi^2/dof$        	& ~  & $2.66/3$ & ~~~\\ [0.25cm]
Probability        	& ~  & 44.6 \% & ~~~\\ [0.20cm]
\hline
$\Gamma(\eta \ra \ggam)$ (keV)  & $0.46 \pm 0.04$  & $0.46 \pm 0.01$  & 0.00 $\sigma$\\ [0.25cm]

$\Gamma(\etp \ra \ggam)$ (keV)  & $4.29 \pm 0.15$  & $4.34 \pm 0.14$  & 0.24  $\sigma$\\ [0.20cm]
\hline

$\Gamma(\eta \ra \pi^+ \pi^-\gamma)$  (eV) & $55 \pm 5$  & $56.64 \pm 1.71$ &  0.31  $\sigma$ \\ [0.25cm]
$\Gamma(\etp \ra \pi^+ \pi^-\gamma)$ (keV) & $60 \pm 5$  & $49.75 \pm 3.88$  & 1.62  $\sigma$\\ [0.20cm]

\hline
\end{tabular}
\caption{
Simultaneous fit of the four HLS anomaly equations (Eqs. (\ref{eq11})) and 
(\ref{eq21}) with only $x$  free (Approximate Field Transformation). 
First data column gives the recommended values \cite{PDG02}.
}
\label{T4}
\end{center}
\end{table}

\subsection{Fit Results With Approximate Field Transformation}

\indent \indent
These equations depend only on $f_\pi$ , $z=[f_K/f_\pi]^2$
and on $x$. In the present framework, $\theta_P$ is no longer an 
independent parameter as it can be algebraically derived from Eq. 
(\ref{eq10}).

One can consider legitimate to still fix $f_\pi$
to its experimental value (92.42 MeV)~; this
is also true for $z$ (see Table \ref{T0}). Therefore,
our set of anomaly equations depends on only  one parameter
$x$, we choosed  previously to fix from fit results
to radiative decays \cite{rad,chpt}. Releasing
the constraint Eq. (\ref{eq10}) would only add a comfortable
(and useless) parameter freedom to the fits 
presented just below.

Therefore, one  considers here Eqs. (\ref{eq11}) and (\ref{eq21})
by themselves and attempt to fit them as a constrained
system of 4 equations with only {\it one} unknown ($x$).
The results are expected to provide consistency with those
obtained for the same parameters and physics
quantities derived elsewhere \cite{rad,chpt} from fit to $VP\gamma$ 
decay modes. 

The $\ggam$ partial widths are related with the amplitudes
given in Eqs. (\ref{eq11}) by~:
\be
\displaystyle \Gamma (X \ra \ggam)=
\frac{M_X^3}{64 \pi} \left | G_X(0) \right|^2
\label{eq30}
\ee

On the other hand, the partial widths
$\Gamma (X \ra \pi^+ \pi^- \gamma )$ given
by Eq. (\ref{eq27}) where the coefficients $A_X$ 
given by Eq. (\ref{eq29}), depend only on $\rho$
properties already derived in \cite{ff1} by a fit to the
pion form factor. The errors on $A_X$ are taken
into account in the fit as they are independent of $x$.

One has performed a fit of these four partial widths
keeping first $z$ fixed and allowing $x$ to vary. 
The results are summarized in Table \ref{T4}.

It is clear that the fit is fairly successfull and
represents the most constrained fit of the four
partial widths ever proposed. One should remark that the best
fit returns a value for $x$ perfectly consistent
with our previous fits to solely  the $VP \gamma$ 
radiative decays, as can be concluded by comparing to
its input value (see Table \ref{T0}). The corresponding 
value for $\theta_P$ is not changed compared to our
previous estimates from fit to $VP\gamma$ decay
modes~: $\theta_P=-10.48^\circ \pm 0.18^\circ$.

We do not give the estimates for derived quantities
($f_0$, $f_8$, $\theta_0$, $\theta_8$) as they
practically coincide with the values given in 
\cite{chpt} and are all in good correspondence with 
expectations. Concerning partial widths, three
out of four reach a significance much better than the
1 $\sigma$ level~; the worst   case is 
$\Gamma(\etp \ra \pi^+ \pi^- \gamma)$ for which the
distance to the recommended value \cite{PDG02} is ``only'' 
$\simeq 1.6 ~\sigma$.

The  fit quality yielded ($\chi^2/dof= 2.66/3$) is such
that releasing  also $z$ can look like an academic exercise.
It has nevertheless been performed as some correlation 
could spoil numerically the connection between $z$ and 
$[f_K/f_\pi]^2$.

The fit returned $x=0.908 \pm 0.021$ and 
$z=1.488 \pm 0.054$ with $\chi^2/dof=2.62/2$,
practically unchanged, corresponding to a 27 \% 
probability (the worse significance is due to having
less degrees of freedom). The correlation coefficient
is $+0.67$, and the minimization does not spoil
the numerical values found elsewhere \cite{rad,chpt}
for the same parameters.
 
  Therefore, this leads us to conclude that 
  the $VP\gamma$ decay modes on the one hand, and
 the four standard anomalous $\eta/\etp$ decay modes  on the 
 other hand, yield  information fairly consistent with each 
 other. This also means that the anomaly equations we derived
 are consistent and that the approximate field transformation
 (leading order in breaking parameters) on which they rely
 match well the present level of accuracy of the data.
 This statement will be confirmed directly shortly.
 
\subsection{$\theta_P$ versus $x$}

\indent \indent
Eq. (\ref{eq10}) corresponds to setting the EChPT decay
constant \cite{Kaiser0,Kaiser} $F_\eta^0$ to zero. This is
rigorously expressed in the broken HLS model by~:
\be
\displaystyle
\tan{\theta_P}= \frac
{\langle0|J_{\mu}^0|\eta^8(q)\rangle}
{\langle0|J_{\mu}^0|\eta^0(q)\rangle}=\frac{b_0}{f_0}
\label{eq31}
\ee
\noindent in term current matrix elements and of
their expressions \cite{chpt}. On the other hand, detailed
computation yields~:
\be
\displaystyle
\frac{f_0}{f_\pi}=\frac{2+z}{3x} ~~~, 
~~~~\frac{b_0}{f_\pi}=\frac{\sqrt{2}}{3}~(1-z)~x
\label{eq32}
\ee

As clear from its expression $f_0$ keeps the first non--leading
contribution in breaking parameters ($x=1+[x-1]$)~; for $b_0$
 one has naturally chosen to replace $x$ by 1 and
this leads to Eq. (\ref{eq10}). However, one may be tempted
to keep it and this leads  to replace $x$ by $x^2$ in
the expression Eq. (\ref{eq10}) for $\tan{\theta_P}$~;
this is nothing but changing the existing term of order 
${\cal O}([z-1][x-1]) \simeq 0.05$. 

We have redone the fit just described with 
this change and yielded a slightly better fit
quality than the previous one 
($\chi^2/dof=2.61/3$). This fit returns also
$x=0.902 \pm 0.017$ ($\simeq 0.5 ~ \sigma$ from
its partner in Table \ref{T4}, or also a one percent change) 
and no change at all for the partial 
widths compared to what is displayed in Table \ref{T4}. 

Therefore the sensitivity in describing data is 
not sharply dependent on non--leading 
contributions in $\tan{\theta_P}$ and using  Eq. (\ref{eq10})
with $x$ or $x^2$ gives undistinguishable results, while
the latter might be prefered.

\subsection{Fit Results With Exact Field Transformation}
  
 \indent \indent
In order to check the sensitivity of the model to
some other details of the broken HLS model, we have also 
attempted fits using the exact field transformation \cite{chpt} 
instead of its leading order approximation (see Eq. (\ref{eq8}))~; 
some details and formulae are given in the Appendix.

The main motivation was to figure out the sensitivity
of the data to the approximation performed on the field
transformation.

In this last series of fits, we have kept $z$ fixed~;
therefore the fitting parameters are $\lambda$ (the
basic Nonet Symmetry breaking parameter, see Eq. (\ref{eq7}))
and $\theta_P$, the later being possibly fixed by 
the constraint $\theta_0=0$. The fit results and 
physics quantities of relevance (ChPT decay constants,
mixing angles and partial widths) are given in Table \ref{T5}.

The first conclusion one can draw from this last Table is
that the fit value for $\theta_0$  departs by less than 1 
$\sigma$ from zero and the consequences of this on 
derived physics quantities is simply negligible.
Stated otherwise, the present data are insensitive to 
releasing the constraint $\theta_0=0$. This constraint
allows to extract a value for $\theta_8 = -18.2^\circ$ 
with a very small statistical error ($\simeq 0.25^\circ$)~;
The value found for $f_0$ and $f_8$ are in the usual ballpark
and nothing noticeable  appears compared to the case when 
the approximate field transform was used \cite{chpt}.

The partial widths are still quite consistent with those
in Table \ref{T4}, showing that the refinements introduced
by the exact transformation have no impact on the extracted
width information for $\eta/\etp \ra \ggam$ and
 $\eta/\etp \ra \pi^+ \pi^- \gamma$. 
 
\vspace{0.3cm} 
 
 One might maybe note that, in all our attempts, we never 
 get a solution with the partial width for 
 $ \etp \ra \pi^+ \pi^- \gamma$ larger than its
 recommended value \cite{PDG02}~; so, the observed 1.6 $\sigma$  
 departure looks like some small systematic 
 effect. This could be due to having neglected
 some unidentified tiny (higher order) contribution~; this
 might also indicate that the recommended value
 is slightly overestimated.
 
 On the other hand, we have also reconsidered the 
 problem of which value for $\eta \ra \ggam$
 should be prefered among the the recommended
 value \cite{PDG02} --recently confirmed by
 a direct measurement of this branching fraction
 \cite{Abegg}--, the $\ggam$ measurements and the
 (single) Primakoff effect measurement. This was done
 already in \cite{chpt}, but with only the
 $\eta/\eta \ra \ggam$ modes. In the present framework
 extended to the $\eta/\etp \ra \pi^+ \pi^- \gamma$ decay modes,
 the conclusion is confirmed~: The recommended value 
 is still clearly preferred~; fit quality indicates
  that it could be slightly smaller (in the direction
  of the Primakoff measurement), but larger values
  (in direction of the  $\ggam$ measurements) are 
  clearly disfavored.

\begin{table}[htb]
\begin{center}
\begin{tabular}{lccc}
~ & PDG 2002  &  Fit Result            &  Fit Result  \\
~ & ~         &  $\theta_0$ free       & $\theta_0=0$ \\[0.3cm]
\hline
$\lambda$        & ~  &          $0.23 \pm 0.06 $         &        $0.21 \pm 0.04 $      \\ [0.3cm]
$\theta_P$       & ~  &   $-10.85^\circ \pm 1.27^\circ$   &  $-11.48^\circ \pm 0.02^\circ$ \\ [0.3cm]
$\chi^2/dof$     & ~  &   2.93/2 &    3.20/3  \\ [0.3cm]
Probability      & ~  &   23.1 \% &    36.5\%\\ [0.3cm]
\hline
$\theta_0$       & ~  &   $-1.01^\circ \pm 1.27^\circ$    &           0 \\ [0.3cm]
$\theta_8$       & ~  &   $-19.37^\circ \pm 1.29^\circ$   &  $-18.16^\circ \pm 0.24^\circ$ \\ [0.3cm]
$f_0$            & ~  &   $1.37 \pm 0.03$     &   $1.36 \pm 0.03$ \\ [0.3cm]
$f_8$            & ~  &   $1.34 \pm 0.01$     &   $1.34 \pm 0.02$ \\ [0.3cm]
\hline
\hline
$\Gamma(\eta \ra \ggam)$ (keV)  & $0.46 \pm 0.04$  & $0.45 \pm 0.03$ & $0.44 \pm 0.01$ \\ [0.3cm]

$\Gamma(\etp \ra \ggam)$ (keV)  & $4.29 \pm 0.15$  & $4.37 \pm 0.23$ & $4.20 \pm 0.17$ \\ [0.3cm]
\hline

$\Gamma(\eta \ra \pi^+ \pi^-\gamma)$  (eV) & $55 \pm 5$  & $55.38 \pm 2.78$ & $55.98 \pm 1.76$\\ [0.3cm]
$\Gamma(\etp \ra \pi^+ \pi^-\gamma)$ (keV) & $60 \pm 5$  & $49.40 \pm 4.85$ & $46.93 \pm 4.00$\\ [0.3cm]

\hline
\end{tabular}
\caption{
Simultaneous fit  of the four HLS anomaly equations modified by using the exact
field transformation. The second data column reports on letting free $\lambda$ and  
$\theta_P$, while in the third data column only $\lambda$ is allowed to vary.
}
\label{T5}
\end{center}
\end{table}

 \vspace{0.5cm}

\section{Summary And Conclusion}
\label{conclusion}

 \indent \indent
The conclusions we get are of various kinds. Therefore,
we prefer segmenting into Subsections.

\subsection*{Experimental Relevance Of The Box Anomaly}

 \indent \indent
 Concerning the analysis of a possible occurence of
 the box anomaly phenomenon in $\eta/\etp$ decays,
 the main results reported in the present paper can be summarized 
 as follows~:
 
 \begin{itemize}
 \item There is a strong evidence in favor of a contact term
 contribution in the $\eta/\etp$ decays to $\pi^+ \pi^- \gamma$.
  All aspects (invariant--mass spectra and partial widths)
  of the $\eta/\etp \ra \pi^+ \pi^- \gamma$ decays can be {\it predicted}
  with fair accuracy using a few pieces of information coming  from
   fits to $VP\gamma$  and $Ve^+e^-$ decay modes in isolation and from
  information coming from  fit to the pion form factor.

  \item
  The needed contact term is numerically at the precise value
  predicted for the box anomaly contribution by the anomalous HLS Lagrangian.
  This plays a crucial role in yielding, without any fit, the
  correct dipion invariant--mass spectra and the correct
  partial widths for both the $\eta$ and $\etp$ mesons.
  
 \item If one lets free the parameters defining the $\rho$ meson
 lineshape in the $\eta/\etp$ spectra, they stay very close to the  values
  expected from (independent) fits to the pion form factor if the
  predicted contact term is switched on.
  
  In contrast, if one removes this from the amplitudes, the
  decription is poor and can only be improved by letting
  the $\rho$ lineshape becoming inconsistent with what is expected
  from fits to the pion form factor.
  
  \item The fit value obtained for the single free parameter
  ($x$, accounting essentially for nonet symmetry breaking)
  indicates indubitably that a global description of
  all $VP\gamma$ modes and of the four $\eta/\etp$  decay modes
  examined here is derived with no additional free parameter. 
   
 \end{itemize}
 
   This leads us to conclude to a strong evidence in favor
  of the occurence of the box anomaly phenomenon in 
   $\eta/\etp  \ra \pi^+ \pi^- \gamma$ decays
  at precisely the level expected from the HLS Model
  and the WZW Lagrangian.
  
 \subsection*{Anomaly Equations And Mixing Angles}

\indent \indent
 On the other hand, we have been led to reexamine the validity of the
 one--angle
 traditional equations giving the amplitudes for $\eta/\etp \ra \ggam$
 and  $\eta/\etp \ra \pi^+ \pi^- \gamma$ at the chiral point, when
 breaking flavor SU(3) and Nonet Symmetries~; resp. the triangle
 and box anomaly equations.
   
  We have found that the broken HLS Model leads to 
  one--angle ($\theta_P$) expressions for the anomaly equations
  which match low energy QCD expectations as expressed by (E)ChPT, 
  but are in deep contradiction with the equations traditionally used. 
  
 Instead of depending on three unconstrained parameters, this set of 
 (four) equations we get depends on only one parameter, 
 closely associated with Nonet  Symmetry breaking (called $x$ or 
 $\lambda$ in the body of the text)~; they also depend on 
 $z=[f_K/f_\pi]^2$ which can hardly be considered as a free 
 parameter. They are proved to be easily fulfilled
 by the relevant $\eta/\etp$ partial widths with fair accuracy.
 
 Relying on the condition $\theta_0=0$, well accepted by the 
  existing data, the broken HLS Model leads to
  an expression of  $\theta_P$ in terms of $z$ and $x$ (or $\lambda$) 
  which can be approximated by a simple formula.  Additionally,
  under the same assumption, an equation leading to 
  $\theta_8 \simeq 2 \theta_P$ can be derived.
 
 These equations have been derived from within the framework of 
 the Hidden Local Symmetry  Model appropriately broken. The 
 phenomenological success of this mechanism implies that the 
 BKY SU(3) Symmetry breaking scheme, supplemented with Nonet 
 Symmetry breaking can be considered as the relevant breaking 
 mechanism.
  
 This extended BKY breaking scheme forces to a field
 transformation which admits a reliable approximation 
 valid at leading order in the breaking parameters
 ($[z-1]$, $[x-1]$). The refinements permitted by the
 exact transformation are found beyond the present
 accuracy of the experimental data. 
 
 \subsection*{Perspectives}

\indent \indent
 At the level of accuracy permitted by the existing data,
 the HLS Model (including its anomalous sector) together
 with the  extended BKY symmetry breaking
 scheme, covers successfully all aspects of the
 experimental data examined so far, certainly 
 up to the $\phi$ mass.  
  
  It would be interesting to have improved data in
  order to check up to which accuracy the HLS
  framework is predictive. For this purpose,
  more and better data on the $\eta/\etp$ sector
  would be welcome.
  
  These could come from tau--charm factories
  (CLEO--C and upgraded BESS) which, running
  at the $J/\psi(1S)$, produce very large
  samples of  $\eta/\etp$ mesons under especially
  clean physics conditions. For instance, in the run
  at the $J/\psi(1S)$ foreseen by CLEO--C
  in 2005 $10^9$ events will be collected. This
  will provide 860\,000 $\eta$ produced opposite in azimuth to
  a single monoenergetic photon and 2\,000\,000
  opposite  to $\omg/\phi$. The corresponding 
  $\etp$ decay modes will provide samples
  of about 4\,300\,000 $\etp$   produced opposite to
  a single photon and  500\,000  opposite to
  $\omg/\phi$. This should allow an exhaustive study
  of the $\eta/\etp$ system
  and a much better understanding of low energy QCD.

\vspace{1.0cm}
\begin{center}
{\bf Acknowledgements}
\end{center}

We gratefully ackowledge V. L. Chernyak (Budker Institute, 
Novosibirsk) for a critical reading of the manuscript and for 
valuable remarks and comments. We also ackowledge G. Shore
(Swansea University, UK) for useful correspondence and reading 
the manuscript. Fermilab is operated by URA under DOE contract No.
DE-AC02-76CH03000. 
  
\newpage
\renewcommand{\theequation}{\Alph {section} . \arabic {equation} }
\setcounter{section}{1}
\setcounter{equation}{0}
\section*{Appendix A}
\subsection*{A1 : The Exact Field Transformation}

\indent \indent
As stated in Section \ref{breaking}, the field transformation
given by Eq. (\ref{eq8}) is an approximation of the full
transformation which has been derived in \cite{chpt}.

In order to bring the kinetic energy part of the U(3)/SU(3)
broken HLS Lagrangian into canonical form (see Eq. (\ref{eq7})), 
it is appropriate to perform the renormalization  in two steps. One first 
diagonalizes the
standard ${\cal L}_{HLS}$ piece using the field transformation
Eq. (\ref{eq6}). This makes the Lagrangian canonical for
the $\pi/K$ sector, and one yields intermediate fields
for the isoscalar sector (double prime fields). 
In terms of bare fields, we have~:
\be
\left [
\begin{array}{ll}
\eta_8^{\prime \prime}\\[0.5cm]
\eta_0^{\prime \prime}
\end{array}
\right ]
= \displaystyle z r
\left [
\begin{array}{ll}
\displaystyle ~~\cos{\beta} & \displaystyle 
\displaystyle ~~~~~~~~~~~-\sin{\beta} \\[0.5cm]
\displaystyle - \sin{\beta}&
\displaystyle ~~ ~~\cos{\beta} -\frac{1}{\sqrt{2}} \sin{\beta}
\end{array}
\right ]
\left [
\begin{array}{ll}
\eta_8\\[0.5cm]
\eta_0
\end{array}
\right ]
\label{A1}
\ee
\noindent where, one has defined~:
\be
r= \frac{\sqrt{(2z+1)^2+2(z-1)^2}}{3z}  \simeq 0.90~~~,
~~~ \tan{\beta}=\sqrt{2}~ \frac{z-1}{2z+1} \simeq 0.20~~~~~.
\label{A2}
\ee
\noindent This transformation brings the kinetic term in the following
form \cite{chpt}~:
\begin{equation}
2 ~{\rm T}=[\partial \eta_8^{\prime \prime}]^2 +
[\partial \eta_0^{\prime \prime}]^2
+ \lambda r~[\sin{\beta} ~\partial \eta_8^{\prime \prime}+ \cos{\beta} ~\partial 
\eta_0^{\prime \prime}]^2.
\label{A3}
\end{equation}
\noindent The transformation to fully renormalized fields (primed fields) 
is performed with~:
\be
\left [
\begin{array}{ll}
\eta_8^{\prime}\\[0.5cm]
\eta_0^{\prime}
\end{array}
\right ]
= \displaystyle 
\left [
\begin{array}{ll}
\displaystyle ~~1+v\sin^2{\beta} & \displaystyle 
\displaystyle ~~~~~~~v\sin{\beta}\cos{\beta} \\[0.5cm]
\displaystyle ~~v\sin{\beta}\cos{\beta}&
\displaystyle ~~ ~~~~~ 1+v\cos^2{\beta}
\end{array}
\right ]
\left [
\begin{array}{ll}
\eta_8^{\prime \prime}\\[0.5cm]
\eta_0^{\prime \prime}
\end{array}
\right ]
\label{A4}
\ee
\noindent where $v$ carries the real information
about Nonet Symmetry breaking (see Eq. (\ref{eq7}))~: 
\be
v=\sqrt{1+\lambda r^2} -1 \simeq 0.10
\label{A5}
\ee
\noindent That the transformation combining Eqs. (\ref{A1}) and
(\ref{A4}) results, at leading order in the breaking parameters
$[z-1]$ and $[x-1]$, into a transformation as simple as Eq. (\ref{eq8}), 
is a little bit unexpected. As noted in the main text, there are several
combinations involving $\lambda$ which are equivalent to $x$
at leading order~; they are all of the form exhibited by Eq. (\ref{eq9})
which is typically  a good representation of $x$ in terms of $\lambda$.

This remainder makes clear why $x$ is influenced by the SU(3)
symmetry breaking. A typical expression for $x$ is~:
\be
\displaystyle x = \frac{1}{\sqrt{1+v}}
\label{A6}
\ee

\subsection*{A2 :  The Anomalous Amplitudes At The Chiral Limit}

\indent \indent 
The expressions for the anomalous amplitudes at the chiral limit,
when using the exact transformation, are easy to get. They amount 
to the following changes for the triangle anomaly expressions
in Eqs. (\ref{eq11})~:
\be
\begin{array}{lll}
\displaystyle ~~~~\frac{5z- 2}{3z} 
        & \displaystyle \Longrightarrow \frac{1}{1+v} \left[
	\frac{5z- 2}{3z}+ \frac{v \cos{\beta}}{rz}
	\right]  & ~~~~~~~~~~~~\displaystyle { \rm (octet)}\\[0.6cm]
	
\displaystyle \sqrt{2} \frac{5z+1}{3z} ~x
        & \displaystyle \Longrightarrow \frac{1}{1+v} \left[
	\sqrt{2}\frac{5z+1}{3z} - \frac{v \sin{\beta}}{rz}
	\right] & ~~~~~~~~~~~~\displaystyle { \rm (singlet)}
\end{array}
\label{A7}
\ee

The octet and singlet combinations for the box anomalies
can easily be identified by the occurence of the $x$ factor
in Eqs. (\ref{eq19}) and (\ref{eq21}). The changes to be performed there
are~:
\be
\begin{array}{lll}
\displaystyle 1 &\displaystyle \Longrightarrow \frac{1}{1+v}\left[
	 1 + \frac{v\cos{\beta}}{rz}
	 \right]  &~~~~~~~~~~~~~ \displaystyle { \rm (octet)}\\[0.6cm]
	 
\displaystyle x &\displaystyle \Longrightarrow \frac{1}{1+v}\left[ 
	 1 -\frac{v\sin{\beta}}{rz \sqrt{2}}
	 \right]  & ~~~~~~~~~~~~\displaystyle { \rm (singlet)}
\end{array}
\label{A8}
\ee

It is worth remarking that the exact field 
transformation changes the 
$(\rho\gamma \eta) $ and $(\rho\gamma \etp)$ 
coupling constants in such a way that the $c_X$'s  
--modified as just stated-- still factor out from
their expression. Therefore, Eqs. (\ref{eq18}) and 
(\ref{eq24}) keep their structure and the
decay invariant--mass spectra  for the $\eta/\etp$ 
are the same as for the approximate field
transformation.

\subsection*{A3 : Decay Constants And Mixing Angles}

\indent \indent
One can express easily the EChPT coupling constants 
($f_0$ and $f_8$) and mixing angles  ($\theta_0$ and 
$\theta_8$) in terms of the parameters mixing
$\lambda$ and $\beta$ defined in the previous 
Subsections.  

The following matrix elements of axials currents 
can be defined in the broken HLS Lagrangian \cite{chpt}~:
\bea\non
\langle0 J_{\mu}^8|\pi^8(q)\rangle=i f_8 q_{\mu}\,,\,\,\,\,&&
\langle0|J_{\mu}^0|\eta^0(q)\rangle=i f_0 q_{\mu}\\
\langle0|J_{\mu}^8|\eta_0(q)\rangle=i b_8 q_{\mu}\,,\,\,\,\,&&
\langle0|J_{\mu}^0|\pi^8(q)\rangle=i b_0 q_{\mu}
\label{A9}
\eea
One can easily write down the currents and their matrix 
elements (see \cite{chpt}, Section 6) in the case when the field 
transformation is not approximated by Eq. (\ref{eq8}). 
Using the notations defined in  \cite{chpt}, one finds first~:
\be
\begin{array}{lll}
\displaystyle \frac{f_8 }{f_\pi}=&  ~~~\displaystyle\frac{rz}{1+v}   \left[
	 1 + v\cos{2\beta}~
	 \right] ~ \cos{\beta}       \\[0.6cm]
	 
\displaystyle \frac{b_8 }{f_\pi}=& \displaystyle -\frac{rz}{1+v}\left[ 
	 1 -v\cos{2\beta}~
	 \right] ~ \sin{\beta}
\end{array}
\label{A10}
\ee
Defining the following parameter combinations~:
\be
\begin{array}{ll}
\displaystyle
h_1=\cos{\beta} - \frac{\sin{\beta}}{\sqrt{2}} \simeq 0.90 ~~~,& ~~
\displaystyle
h_2=\lambda \cos{\beta}~ - \frac{1+\lambda}{\sqrt{2}} \sin{\beta} \simeq 0.03
\end{array} 
\label{A11}
\ee
\noindent one also yields~:
\be
\begin{array}{lll}
\displaystyle \frac{f_0 }{f_\pi}=&  ~~~\displaystyle
\frac{rz}{1+v}   \left[~
         h_1 (1+\lambda)  +   v  (2 \cos{\beta} +h_2) \sin^2{\beta}~
	 \right]       \\[0.5cm]
	 
\displaystyle \frac{b_0 }{f_\pi}=& \displaystyle -\sin{\beta} 
	\frac{rz}{1+v}   \left[~
		(1+v\cos{2\beta}) -h_1 v  \cos{\beta}~
	 ~\right]       
\end{array}
\label{A12}
\ee

\subsection*{A4 : The Condition $\theta_0=0$}

\indent \indent 
Phenomenology \cite{chpt} as well as explicit EChPT
computations \cite{Holstein4} indicate that the
mixing angle $\theta_0$ is very close to zero.
Table \ref{T5} clearly illustrates that present
data are statistically insensitive to letting
$\theta_0$ departing from zero. Under such
conditions, several interesting relations show up.

The definition of the angles $\theta_0$ and $\theta_8$
can be expressed in terms of the parameters in Eq. (\ref{A9})
\cite{chpt}. Using Eq. (\ref{A12}), one can  derive~:
\be
\begin{array}{ll}
\displaystyle
\tan{\theta_8} = \tan{(\theta_P+\varphi_8)}~~~,~~& \tan{\theta_0} = -\tan{(\theta_P-\varphi_0)}
\end{array}
\label{A13}
\ee
\noindent where $ \tan{\varphi_8}=b_8/f_8$ and $ \tan{\varphi_0}=b_0/f_0$
can be explicitly computed. The condition $\theta_0=0$ strictly implies
that $\theta_P=\varphi_0$ which gives~:
\be
\displaystyle
\tan{\theta_P}  = - \frac{1}{1+\lambda}
\left [ \frac{\tan{\beta}}{1 - \frac{1}{\sqrt{2}}\tan{\beta}} \right ]~~
\left [ 1+ \frac{v \tan{\beta}}{\sqrt{2}}  + \cdots~\right ]
\label{A14}
\ee

{From} Eq. (\ref{A6}), the first term can be interpreted as $x^2$
and the product of the first two factors is just Eq. (\ref{eq10})
for $\tan{\theta_P}$ modified with $x^2$. With the values for
$v$ and $\tan{\beta}$ we have mentioned, the leading correction 
amounts to only $1.5 ~ 10^{-2}$. If one keeps a $1/\sqrt{1+\lambda}$
(corresponding to having $x$ in Eq. (\ref{eq10}))
in front of this expression, the correction term gets a additional
contribution $-\lambda/2 \simeq 5. ~ 10^{-2}$  which becomes dominant.
Therefore~: 
\be
\displaystyle \tan{\theta_P}=\sqrt{2} \frac{(1-z)}{2+z} ~x^2
\label{A15}
\ee
\noindent could indeed be prefered to Eq. (\ref{eq10}).

The second information which follows from $\theta_0=0$ is
an approximate relation between $\theta_8$ and the wave--function
mixing angle $\theta_P$~:
\be
\displaystyle
\tan{\theta_8}=2\tan{\theta_P} \left [
1-\frac{\tan{\beta}}{2\sqrt{2}}
+ \cdots ~\right ] 
\label{A16}
\ee
\noindent where the leading correction is $\simeq 7. ~ 10^{-2}$.

\newpage

\newpage

\newpage

\newpage

\newpage

\newpage

\end{document}